\documentclass{aa}  
\usepackage{listings}
\usepackage{graphicx}
\usepackage{txfonts}
\usepackage[multiple]{footmisc}
\usepackage{pdflscape}
\begin{document}

   \title{The most massive, passive, and oldest galaxies at $0.5<z<2.1$: Downsizing signature from galaxies selected from Mg$_{UV}$ index\thanks{Based on data obtained with the European Southern Observatory Very   Large   Telescope,   Paranal,   Chile,   under   Large   Programmes 070.A-9007  and  177.A-0837.  Based  on  observations  obtained  with MegaPrime/MegaCam,  a  joint  project  of  CFHT  and  CEA/DAPNIA,at the Canada-France-Hawaii Telescope (CFHT) which is operated by the National Research Council (NRC) of Canada, the Institut National des  Sciences  de  l’Univers  of  the  Centre  National  de  la  Recherche
Scientifique  (CNRS)  of  France,  and  the  University  of  Hawaii.  This work is based in part on data products produced at TERAPIX and the
Canadian Astronomy Data Centre as part of the CFHT Legacy Survey, a collaborative project of NRC and CNRS.}}
  \titlerunning{Massive galaxies selected by their Mg$_{UV}$  feature}
   \author{R.Thomas
          \inst{\ref{ESOChile}}
           \and  O. Le F\`evre\inst{\ref{LAM}}
           \and G. Zamorani\inst{\ref{BolognaObs}}
           \and B. C. Lemaux\inst{\ref{Davis}}
           \and P. Hibon\inst{\ref{ESOChile}}
           \and A. Koekemoer\inst{\ref{STSI}}
           \and N. Hathi\inst{\ref{STSI}}
           \and D.~Maccagni\inst{\ref{Milano}} 
           \and P.~Cassata\inst{\ref{Padova}} 
           \and L. P. Cassar\`a\inst{\ref{Milano}}
           \and S. Bardelli\inst{\ref{BolognaObs}}
           \and M. Talia\inst{\ref{BolognaObs}}
           \and E. Zucca\inst{\ref{BolognaObs}} }

\institute{European Southern Observatory, Av. Alonso de C\'ordova 3107, Vitacura, Santiago, Chile \label{ESOChile}, \\
              \email{rthomas@eso.org}
\and Aix Marseille Universit\'e, CNRS, LAM (Laboratoire d'Astrophysique de Marseille) UMR 7326, 13388, Marseille, France\label{LAM}
\and INAF - Osservatorio di Astrofisica e Scienza dello Spazio di Bologna, via Gobetti 93/3, I-40129, Bologna, Italy\label{BolognaObs}                
\and Department of Physics, University of California, Davis, One Shields Ave., Davis, CA 95616, USA\label{Davis} 
\and Space Telescope Science Institute, 3700 San Martin Drive, Baltimore, MD, 21218, USA\label{STSI}
\and INAF-IASF Milano, via A. Corti 12I, 20133 Milano, Italy\label{Milano}
\and University of Padova, Department of Physics and Astronomy Vicolo Osservatorio 3, 35122, Padova, Italy\label{Padova}
             }

   \date{Submitted ---; Accepted ---}

 
  \abstract
   {}
   {We seek is to identify old and massive galaxies at 0.5<z<2.1 on the basis of the magnesium index Mg$_{UV}$ and then study their physical properties.}
   {We computed the Mg$_{UV}$ index based on the best spectral fitting template of $\sim$3700 galaxies using data from the VLT VIMOS Deep Survey (VVDS) and VIMOS Ultra Deep Survey (VUDS) galaxy redshift surveys. Based on galaxies with the largest signal to noise and the best fit spectra we selected 103 objects with the highest spectral Mg$_{UV}$ signature. We performed an independent fit of the photometric data of these galaxies and computed their stellar masses, star formation rates, extinction by dust and age, and we related these quantities to the Mg$_{UV}$ index.}
   {We find that the Mg$_{UV}$ index is a suitable tracer of early-type galaxies at an advanced stage of evolution. Selecting galaxies with the highest Mg$_{UV}$ index allows us to choose the most massive, passive, and oldest galaxies at any epoch. The formation epoch t$_f$ computed from the fitted age as a function of the total mass in stars supports the downsizing formation paradigm in which galaxies with the highest mass formed most of their stars at an earlier epoch.} 
   {}


   \maketitle
%

\section{Introduction}

The theory of galaxy formation and evolution is based on the hierarchical model of structure formation \citep{white78}. 
In this model small structures form first and merge together to create bigger systems. These systems then merge with other systems to form even bigger structures. This type of evolution is classically represented by a merger tree (Lacey and Cole, 1993). Intriguingly, various studies have reported that galaxy evolution follows a downsizing pattern \citep{Cowie96,cimatti06,thomas10}. In this paradigm, the most massive galaxies form rapidly at earlier cosmic epochs than the less massive galaxies. In addition, a number of studies have shown that passive galaxies seen today have been formed from an intense star forming event at earlier cosmic times followed by a passive evolution \citep{thomas05}. While this might appear, at first, as an anti-hierarchical behavior, it has been shown that this  can be compatible with the hierarchical scenario, arising in the clustering processes of dark matter halos, provided the physics of baryonic matter is correctly modeled \citep{sparre15, Qu17}. 

The transition from strongly star forming galaxies to passive systems remains to be understood, and therefore we need to identify galaxy samples at intermediate epochs and study their properties to understand how this transition happens. However, it is crucial to define the most suitable type of galaxies to improve our understanding of   the galaxy evolution scenario.
Based on color-magnitude diagrams (hereafter CMD), for example U-V versus M$_{V}$ \citep{Bell04}, galaxies can be separated into two different categories.
Late-type galaxies are known as relatively young low-mass galaxies with active star formation \citep{Brinchmann04, Amorin17,Lumbreras18}. Their spectral light is dominated by the emission of young stars in the UV. These galaxies form what we call the blue cloud in the CMDs. On the other hand early-type galaxies, which form the red sequence in the CMD, are generally seen as old massive galaxies experiencing very little or no ongoing star formation over the last several gigayear of their life \citep{Brian12, Estrada2019}. Because they are thought to undergo limited star formation activity over a large fraction of their existence, these galaxies are also known as passively evolving galaxies (hereafter PEGS). The light emitted by these galaxies is dominated by the oldest population that radiates the most in the near-infrared (NIR). These galaxies have been used to place constraints on galaxy evolution scenarios and in particular the aforementioned downsizing evolution of galaxies at different redshifts (e.g., \citealt{Fritz14, Siudek17}). 


These studies rely on the identification and selection of passive galaxies. Several methods have been used to find such galaxies at low and high redshift using photometry or spectroscopy. Based on photometric data, the classical way of selecting passive galaxies makes use of CMD or color-color diagrams in which galaxies with no (or little) ongoing star formation are located in particular regions of these diagrams. Using rest-frame colors, the UVJ diagram (\citealt{Will09}, \citealt{Bram11}, \citealt{Straatman14}, \citealt{Merlin18}) and the NUVrJ diagram (\citealt{Ilbert13}, \citealt{Davi17}) have been widely used in the literature up to z$\sim$2. Using observed frame colors at higher redshift, the BzK diagram has also been proposed to select this type of galaxies (\citealt{D04}, \citealt{D05}, \citealt{Ono12}).
The efficiency of these methods in separating red galaxies from the rest of the population have been studied through a wide range of redshift domains. 
For the UVJ diagram, the contamination by star forming galaxies is <1\% between z=0.5 and z=2.5, while the completeness of the selection is decreasing with redshift from 97\% at 0.5<z<1.0 to 81\% at 2<z<2.5 \citep{Gu18}.  For the NUVrJ diagram, \citet{Ilbert13} estimated that the completeness decreases slightly with redshift ranging from 95\% at 0.2<z<0.7 to 87\% at 2<z<3. The contamination on the contrary increases with redshift, as 10\% of the non-red galaxies enter the selection area at z<0.7 and reach 60\% at 2<z<3.
Photometric data associated with the widely used spectral energy distribution (SED) fitting technique have also been used in the selection of passive galaxies. From the estimation of the stellar mass and star formation rates (SFR), it is possible to select passive galaxies using the specific star formation rate (hereafter sSFR, defined as the ratio of the SFR to stellar mass). The threshold in sSFR is slightly different from one study to another. In \cite{Cass10}, \cite{ilbert2010}, and \cite{Tamb14}, the threshold to select passive galaxies is set at sSFR$<10^{-2}$Gyr$^{-1}$, while in \cite{McLure12} galaxies are considered passive when sSFR$<10^{-1}$Gyr$^{-1}$. \cite{Ilbert13} found that this selection based on the sSFR is equivalent to that based on the NUVrJ diagram but the classification from the sSFR is more conservative at high redshift.

From a spectroscopic point of view the absence of ongoing star formation can be characterized using emission lines such as [OII] and H$\alpha$, although their presence does not necessarily indicate ongoing star formation (e.g., \citealt{Yan06, Brian10}). The study of such lines, which are tightly connected to star formation, allows for the identification of galaxies with weak ongoing star formation. For instance, \cite{M12} chose galaxies with EW([OII]) and EW(H${\alpha}$), which are both lower than 5\AA,~to select PEGS. In addition to emission lines, continuum features are also of great interest to select galaxies with old stellar populations. Two main spectral indicators have been considered. The so-called $D_{4000}$ break, defined as the ratio of the mean flux redward to blueward of 4000\AA~in 100\AA~wide bandpasses \citep{Balogh99}. 
This break is commonly used as a sign of an already evolved stellar population and, under some hypotheses of  star formation history (SFH) and metallicity, was even used to compute galaxy ages \citep{M12}. This index is the result of the accumulation of metallic absorption lines creating a jump in the spectral continuum. For an observed window between 3500 and 9500\AA~, the $D_{4000}$ break is visible from $z=0$ to $z\sim1.25$.
At higher redshift, where the $D_{4000}$ break is no longer visible in such a wavelength window, another spectral index has been proposed to identify such galaxies: the Mg$_{UV}$ index (\citealt{D05}, hereafter D05). Its presence is the result of the combination of absorption lines such as Mg I, Mg II, and Fe II. 
Even if it is a fainter index, the Mg$_{UV}$ index is a good alternative to the D$_{4000}$ break for passive galaxy selection for two main reasons. For a spectrograph spanning from 3500\AA~ to 9500\AA~ the Mg$_{UV}$ index is available from z$\sim$0.5 to z$\sim$2.25, therefore reaching higher redshift than allowed from the $D_{4000}$ index. Moreover, young, dust-reddened galaxies, which often mimic the colors of a passively evolving stellar population, show low values of the Mg$_{UV}$ \citep{D05}. For these reasons, this index is preferred for the study of  high-redshift galaxy samples for which the fraction of young galaxies can be high.

In this paper we aim at studying high redshift massive and passive galaxies selected using the Mg$_{UV}$ index. The paper is organized as follows. In Section \ref{Data} we present the VIMOS Very Large Telescope Deep Survey
 (VVDS) and the VIMOS Ultra-Deep Survey (VUDS) spectroscopic surveys on which we base our sample selection and analysis. In Section \ref{secsel} we select galaxies based on the measurement of the Mg$_{UV}$ index. In section  \ref{secmasssfr} and \ref{secage} we analyze the evolution of this index against key quantities such as stellar mass and galactic ages. In Section \ref{down} we study the evolution of the formation epoch of our selected galaxies with the stellar mass. 
\section{Data}
\label{Data}
In this section we present both the spectroscopic and photometric data we used in this paper. These data come from the publicly available Deep and Ultra-Deep samples of the VVDS (\citealt{OLF4}, \citealt{OLF05}, \citealt{OLF13}) and from the VUDS (\citealt{OLF15}). We give a brief description of the surveys and the available data in this section.

The VVDS is a magnitude selected spectroscopic redshift survey carried out on the VIMOS spectrograph installed at the Nasmyth B focus of the Very Large Telescope, Chile \citep{OLF03}. This survey targeted the CFHTLS-D1 area of the XMM-Large Scale Structure survey (XMM-LSS) field and is composed of three main parts. We used the VVDS Deep, which is composed of $\sim$11500 galaxies in a region of 0.74 deg$^{2}$ and the VVDS Ultra-Deep that contains $\sim$1000 galaxies down to \textit{i}$_{AB}$ = 24.5 on an area of 500 arcmin$^{2}$. The deep sample of the VVDS used the low resolution spectrograph LRRED covering the range 5500$\leq\lambda\leq$9350~\AA. 
This permits us to observe important spectral features like the [OII]3727\AA~emission line from z$\sim$0.5 to z$\sim$1.5. For the ultra-deep sample, both blue and red gratings (LRBLUE and LRRED) were used to cover 3650$\leq\lambda\leq$9350~\AA~. The data reduction was carried out using the VIPGI software \citep{Sco05}, while the redshift measurements were performed with the EZ software \citep{Gari10}. Each redshift measurement is accompanied by a quality flag indicating the probability of the redshift to be correct. The flag system consists of six different flags. Flags 2, 3, 4, and 9 (for objects with a single emission line) are the most reliable flags with a probability to be correct of 75\%, 95\%, 100\%, and 80\%, respectively. A quality flag of 1 indicates a probability of being correct of 50\%, while a quality flags of 0 indicates that no redshift could be assigned.
The VVDS catalog has been matched to existing multiwavelength photometric catalogs. The catalogs we use in this paper contain the  Canada France Hawaii Telescope Survey (CFHTLS; T0005 release) \textit{ugriz} bands; these data were obtained with the Megacam camera. WIRCAM-\textit{JHKs} data are also available from the WIRDS survey (WIRDS; \citealt{Bielby12}).

We also used VUDS, a spectroscopic galaxy survey aimed at studying galaxy evolution in the redshift range $2<z<6+$, that has targeted $\sim$10000 objects. Galaxies have been selected in three widely observed fields to mitigate cosmic variance: COSMOS, ECDFS, and VVDS-2h fields. For 90\% of the survey, the selection was carried out with the photometric redshift method using the LePhare software \citep{Ilbert06}. For the remaining 10\%, color-color Lyman break selection     was used (\textit{ugr}, \textit{gri,} and \textit{riz} diagrams). As for the VVDS, VUDS was carried out with the VIMOS spectrograph at the VLT and each target was observed for $\sim$14 h in the wavelength range 3500$\leq\lambda\leq$9350~\AA~at low resolution. Data were reduced with the VIPGI software and spectroscopic redshifts measured with the EZ tool. The reliability flag system is equivalent to that described above for the VVDS survey. 
The ECDFS field contains the \textit{U,B,V,R,I,Z,J,H,K} bands and IRAC channels from the catalog assembled by \cite{Cardamone10} in the MUSYC survey. The COSMOS fields come with  \textit{u*} band data from the CFHT, Subaru imaging (\textit{B, V, g+, r+, i+, z+}) in the optical and Ultra Vista (\textit{J, Ks}) ,and IRAC (two first channels) for the IR. Finally, for the VVDS-2h field observed in VUDS, we used\textit{ u, g, r, i, z} observations, which are available from the CFHTLS with Megacam down to \textit{i}$_{AB}$ = 25.44 at 50\% completeness using the data release 6 \citep{Cuillandre12}. In the  NIR domain we used YJHK bands obtained with WIRCAM at CFHT down to Ks$_{AB}$ = 24.8 also at 50\% completeness \citep{Bielby12}.\\
 
\section{Galaxy selection}
\label{secsel}

\subsection{Spectroscopic corrections}
\label{spec_res_sec}
Given its wavelength limits the Mg$_{UV}$ index can be measured in VVDS Deep sample between z$\sim$1.29 and z$\sim$2.25 while it is present in the VVDS Ultra-Deep and VUDS samples between z$\sim$0.5 and z$\sim$2.25. Considering all redshift flags (see discussion in Sect.\ref{flags}) we measured Mg$_{UV}$ in 3711 galaxies. The repartition among the different samples is presented in Table \ref{Table_avail}.

\begin{table}[h!]
\centering
\caption{Selection of our sample of UV-selected galaxies. The top part of the table shows the availability of the Mg$_{UV}$ in each galaxy survey and the number of available galaxies. The bottom part presents the selection of the galaxies in the three redshift bins of interest. The value N$_{S/N}$ gives the number of galaxies after the S/N cut and N$_{final}$ gives the final number of selected galaxies, with the  number in parenthesis giving the number of candidates.}
\begin{tabular}{ccccc}
\hline
\multicolumn{4}{c}{Original sample} \\
\hline 
 & \multicolumn{2}{c}{Mg$_{\mathrm{UV}}$} &\multicolumn{2}{c}{N$_{\mathrm{gal}}$}  \\ 
\hline 
\hline
Deep  & \multicolumn{2}{c}{1.29<z<2.25}& \multicolumn{2}{c}{ 950}  \\ 
Udeep & \multicolumn{2}{c}{0.5<z<2.25} &  \multicolumn{2}{c}{ 703}  \\  
VUDS  & \multicolumn{2}{c}{0.5<z<2.25} &  \multicolumn{2}{c}{2058}  \\ 
All   & \multicolumn{2}{c}{All}        & \multicolumn{2}{c}{3711} \\ 

\hline
\hline
\multicolumn{4}{c}{Sample selection} \\
\hline 
Redshift & N$_{\mathrm{Tot}}$  & N$_{\mathrm{S/N}}$ & N$_{\mathrm{final}}$   \\ 
\hline 
\hline
0.5<z$\leq$0.9 &  800 & 571 & 27 (24+3) &\\ 
0.9<z$\leq$1.9 &  2075 & 1414 & 37 (29+8) \\  
1.9<z$\leq$2.25 & 836 & 674 & 39 (22+17) \\ 
All & 3711 & 2659 & 103 (75+287) \\ 
\hline
 
\end{tabular} 
\label{Table_avail}
\end{table}

As we used spectroscopic data, the galaxy spectra we used must be corrected for any instrumental signature that may affect the measurements. As mentioned in Section \ref{Data}, spectra have been  processed following standard and rigorous methods. Nevertheless, some artifacts and residuals may still be present in the spectra. To tackle this problem we performed a spectral fit of our sample of 3711 galaxies via the new SPARTAN software (Thomas et al, in preparation, see Appendix \ref{SPARTAN} for a brief overview). The fit was performed considering a wide parameter space. We used  physical conditions when building stellar population models. This parameter space is summarized in Table \ref{Param1}.

\begin{table}[h!]
\centering
\caption{Template library used for the spectroscopic correction fitting.}
\begin{tabular}{cc}
\hline
Parameter name & Range \\ 
\hline 
\hline
SSP models  & BC03 \\ 
IMF & Chabrier (2003) \\  
Metallicity & 0.02Z$_{\odot}$ < Z < 2.5Z$_{\odot}$ \\ 
Star formation history (SFH) & Exponentially delayed\\
SFH timescale [Gyr] & 0.1 to 5.0 Gyr \\
Age [Gyr] & 0.05 up to Age$_{U}$(z)\\
Dust attenuation & Calzetti\\ 
& 0.0<E(B-V)<0.6\\
IGM & free parameter (z>1.5 only)\\
\hline 
\end{tabular} 
\label{Param1}
\end{table}

We used BC03 models \citep{BC03} with a Chabrier initial mass function (hereafter IMF; \citealt{Chab03}). The assumed SFH is a delayed exponential with a timescale parameter, $\tau$, ranging from 0.1 Gyr to 5.0 Gyr. We used the Calzetti's dust extinction prescription \citep{Calzetti00} with values from E(B-V)$_{s}$=0.0 to E(B-V)$_{s}$=0.6. For galaxies with $z>1.5$ we used a free intergalactic medium (IGM) extinction prescription from \cite{Thomas17}, while it could be neglected for galaxies at lower redshifts. The stellar-phase metallicity we used (hereafter referred simply as metallicity) ranges from 0.02Z$_{\odot}$ to 2.5Z$_{\odot}$, where Z$_{\odot}$  is the solar metallicity. Finally, we allowed the ages to vary from 0.05 Gyr to 13.5 Gyr. For a given galaxy at a given redshift, this range of age is not allowed to go over the age of the Universe at the considered redshift. The redshift at which we fit our galaxies is the spectroscopic redshift measured during the data processing of each survey. It is worth noting that since we used all non-zero redshift flags, we could allow for the redshift to vary during the fitting process to account for small possible variations that are difficult to estimate during the redshift measurement, 
especially in the case of flag=1 redshifts. We chose not to take this into account as the goal of our selection is to identify clear Mg$_{UV}$ signatures rather than building a complete sample.

\begin{figure}[h!]
\centering
\includegraphics[width=0.5\textwidth]{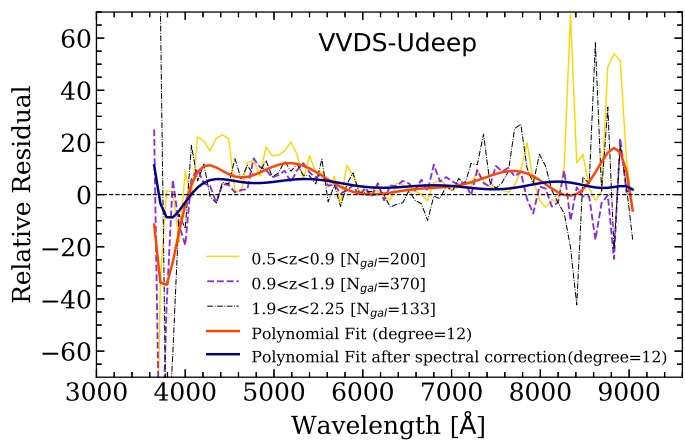}
\caption{Median spectral residual estimated from the fit of our selection of galaxies with SPARTAN in the VVDS Ultra-Deep survey. We show the relative residual in three redshift bins: 0.5<z$\leq$0.9 (yellow line), 0.9<z$\leq$1.9 (purple dashed line), and 1.9<z$\leq$2.25 (black dot-dashed line). We indicate the number of fits used in the estimation of the residual for each redshift bin. We also show the polynomial fit of the average residual before (orange thick line) and after (blue thick line) spectral correction. The fit is performed on the average of all the redshift bins and enters in the correction of the spectroscopy.}
\label{residual}
\end{figure}

From each spectral fit we computed the relative spectral residual, $R_{r}(\lambda)$, with respect to the best fit template. It is given by
\begin{equation}
R_{r}(\lambda) = \frac{F_{obs}(\lambda) - F_{Best\;Fit}(\lambda)}{F_{Best\;Fit}(\lambda)},
\label{res}
\end{equation}
where $F_{obs}$ and $F_{temp}$ are the observed flux and the flux from the best fit template. We computed the median residual in three different redshift bins: $0.5<z<0.9$, $0.9<z<1.9,$ and $1.9<z<2.25$. These global residuals were computed by taking, at a given wavelength, the median of all the residual at this wavelength. This median residual, which is taken as an indication of possible systematic shifts in the observed fluxes as a function of wavelength, is shown in Fig.\ref{residual} for the VVDS Ultra-Deep sample and for each redshift bin. 

Fig.\ref{residual} shows that the residual is on average $\sim$10\%; it is similar for each field. It is particularly interesting to observe that this residual does not evolve with redshift.
This is explained by the fact that the observed-frame window corresponds to a different rest-frame window for each galaxy. Therefore, when averaging in a given redshift bin the effect of the individual redshift is washed out. Thus, the behavior of the residual with wavelength that we computed is the result of either instrumental, observational, or reduction effects. \\
The fitting residuals exhibits strong peaks and troughs at $\lambda>8000$\AA~for galaxies observed at all redshifts. These features indicate large excesses or deficits of flux in the data relative to the expectations from the models. This part of the spectral region, not coincidentally, is where the density of airglow lines is the highest and also where fringing for VIMOS is at its worst. Consequently, it is likely that these features come from issues related to over- or under-subtraction of the airglow lines during the reduction process. At the other edge of the wavelength window, we see that the residual significantly decreases below 4100\AA. This implies that a significant amount of flux is missing in the bluer part of the spectra. This effect could be due to the atmospheric refraction that spreads the light before entering the telescope and which is maximal in the blue or from poor response correction. Since the residuals are equivalent in each redshift bin, we average them and fit a polynomial function (with a degree 12). This fit is shown in Fig.\ref{residual} as well. This fit is used to correct the science spectra to correct for instrumental, observational, and reduction effects. It is worth mentioning that to conserve the S/N we also applied the same corrections to the error spectra. We applied this method for all the data presented in the previous section. After these corrections the averaged relative residual goes down to below 3.5\% (see Fig.\ref{residual}).

\subsection{Signal to noise and redshift flags}
\label{flags}
The measurement of spectral indexes is strongly affected by the signal-to-noise ratio (hereafter S/N)\ of the spectra. Therefore, we compute the S/N of our 3711 galaxies using the prescription of \citet{Stoehr08}, which assumes that the flux in two resolution elements apart is not correlated and that the noise is normally distributed. To compute the S/NR, we chose feature-free spectral regions. For galaxies at 0.5<z<0.9, 0.9<z<1.9, and 1.9<z<2.25 we computed the S/N in the following rest-frame regions: 4360<$\lambda$<4560, 2950<$\lambda$<3150, and 2000<$\lambda$<2220, respectively.  In the spirit of keeping as many galaxies as possible in our sample we keep galaxies with a S/N per resolution element higher than 2. This leads to a S/N-selected sample of 2659 galaxies.\\
As presented in section \ref{Data}, the surveys we used come with a redshift flag system that assesses the reliability of the redshift measurement, which is not directly related to the data quality or the S/N. As we considered galaxies in a redshift range where the lack of strong spectral indices makes the redshift measurement particularly difficult (especially between z$\sim$1.5 and z$\sim$2.0), we do not include a criterion based on the redshift flag in our selection. We discuss this aspect of our approach in the next subsection.

\subsection{Measurements of Mg$_{UV}$}
\begin{figure}[h!]
\centering
\includegraphics[width=0.5\textwidth]{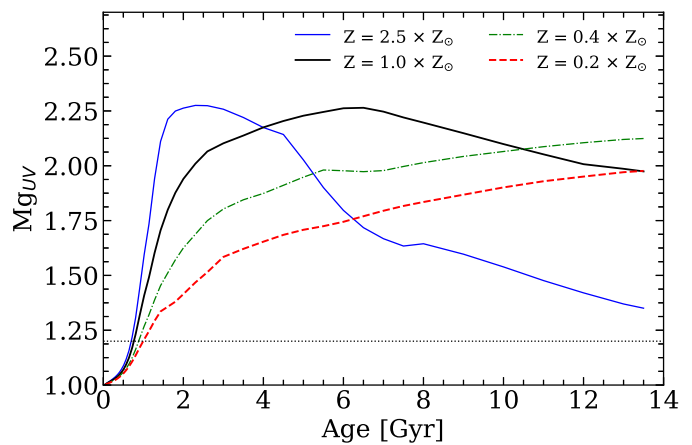}
\caption{Evolution of the strength of the Mg$_{UV}$ index as function of time for four metallicities: Z$_{\odot}$ (solar, black solid line), 0.2$\times$Z$_{\odot}$ (red dashed line), 0.4$\times$Z$_{\odot}$ (green dot-dashed line), and 2.5$\times$Z$_{\odot}$ (solid blue line). The index is computed from SEDs created from the BC03 models \citep{BC03} with an exponentially delayed SFH with an e-folding time of 0.1 Gyr and no dust extinction. The index was computed directly on the synthetic spectra. }
\label{mguv_ev}
\end{figure}

As shown by \cite{D05}, the Mg$_{UV}$ index is an age-sensitive index that already appears in evolved galaxies. This index is defined as the combination of three 100\AA~wide bandpasses in the UV domain of the rest-frame spectrum written as
\begin{equation}
Mg_{UV}= \frac{2\int_{2625}^{2725} f_{\lambda} d\lambda}{\int_{2525}^{2625} f_{\lambda} d\lambda+\int_{2725}^{2825} f_{\lambda} d\lambda}.
\label{mguv_def}
\end{equation}
Fig.\ref{mguv_ev} shows the evolution of the Mg$_{UV}$ strength as a function of age for four metallicities in the case of a galaxy with an exponentially delayed SFH and no extinction. 
This figure shows that the strength of this index rises rapidly for galaxies between 0.5 Gyr and 2 Gyr for every metallicity. We note that a faster rise occurs for galaxies at higher metallicities. For galaxies with older ages, the evolution of the index is strongly dependent on the metallicity. For subsolar metallicities the strength of the index continues to rise up to old ages. For solar and super-solar metallicities, the evolution of Mg$_{UV}$ reaches a maximum and then decreases toward old ages. For solar metallicity the peak is at $\sim$6.5 Gyr while for super-solar metallicity this peak happens earlier in the evolution at $\sim$2 Gyr. 
To create our final sample we computed the Mg$_{UV}$ index for the spectra and the best fit template. In order to keep as many galaxies as possible, we kept in our sample all the galaxies with a Mg$_{UV}$ index of at least 1.1 in at least one of the two measurements. Based on the evolution given in Fig.\ref{mguv_ev}, this ensures to select galaxies with an age of at least $\sim$0.5 Gyr. We then inspected all the candidates by eye and generate two subsamples. 

\begin{figure}[h!]
\centering
\includegraphics[width=8.cm]{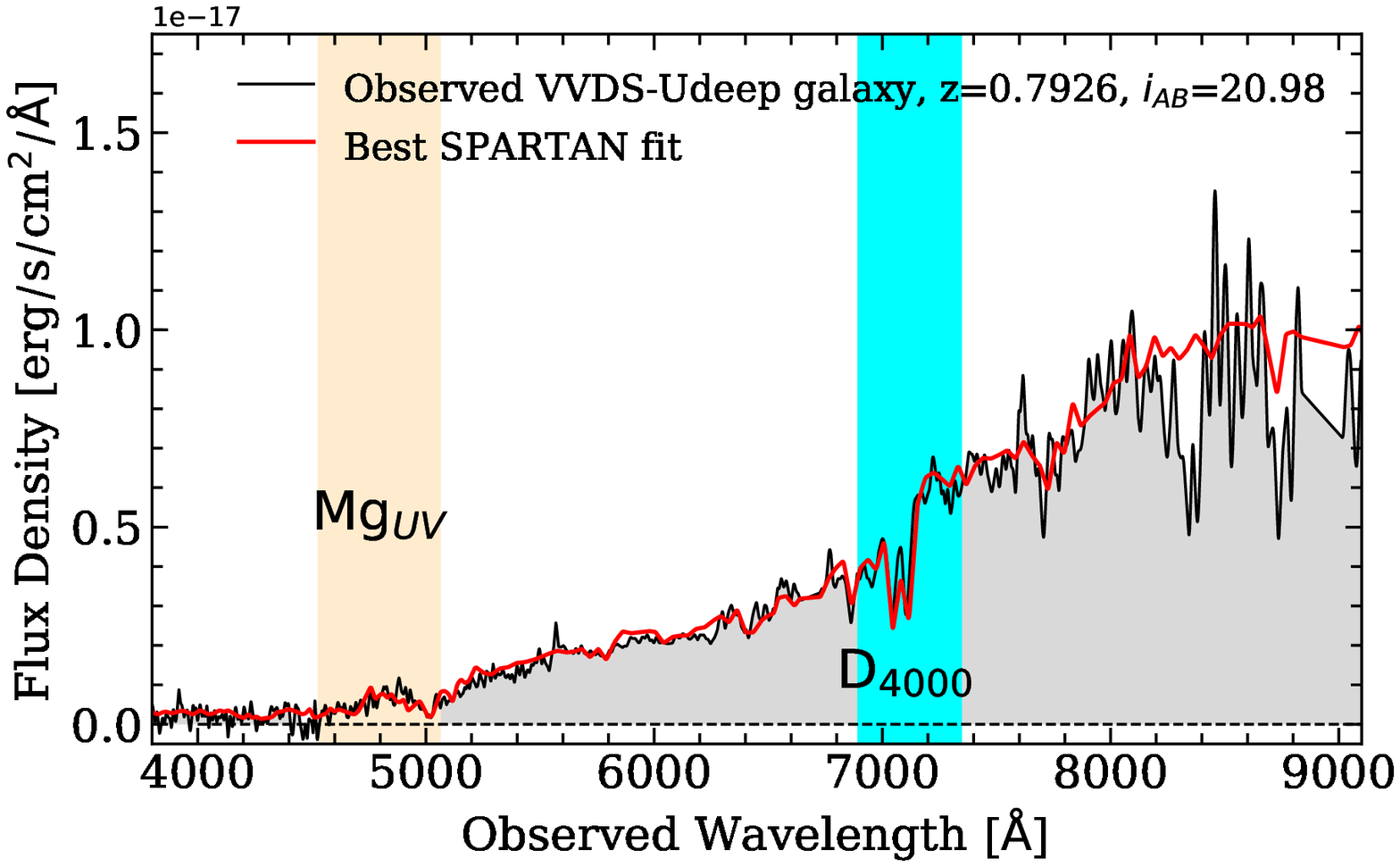}
\includegraphics[width=8.cm]{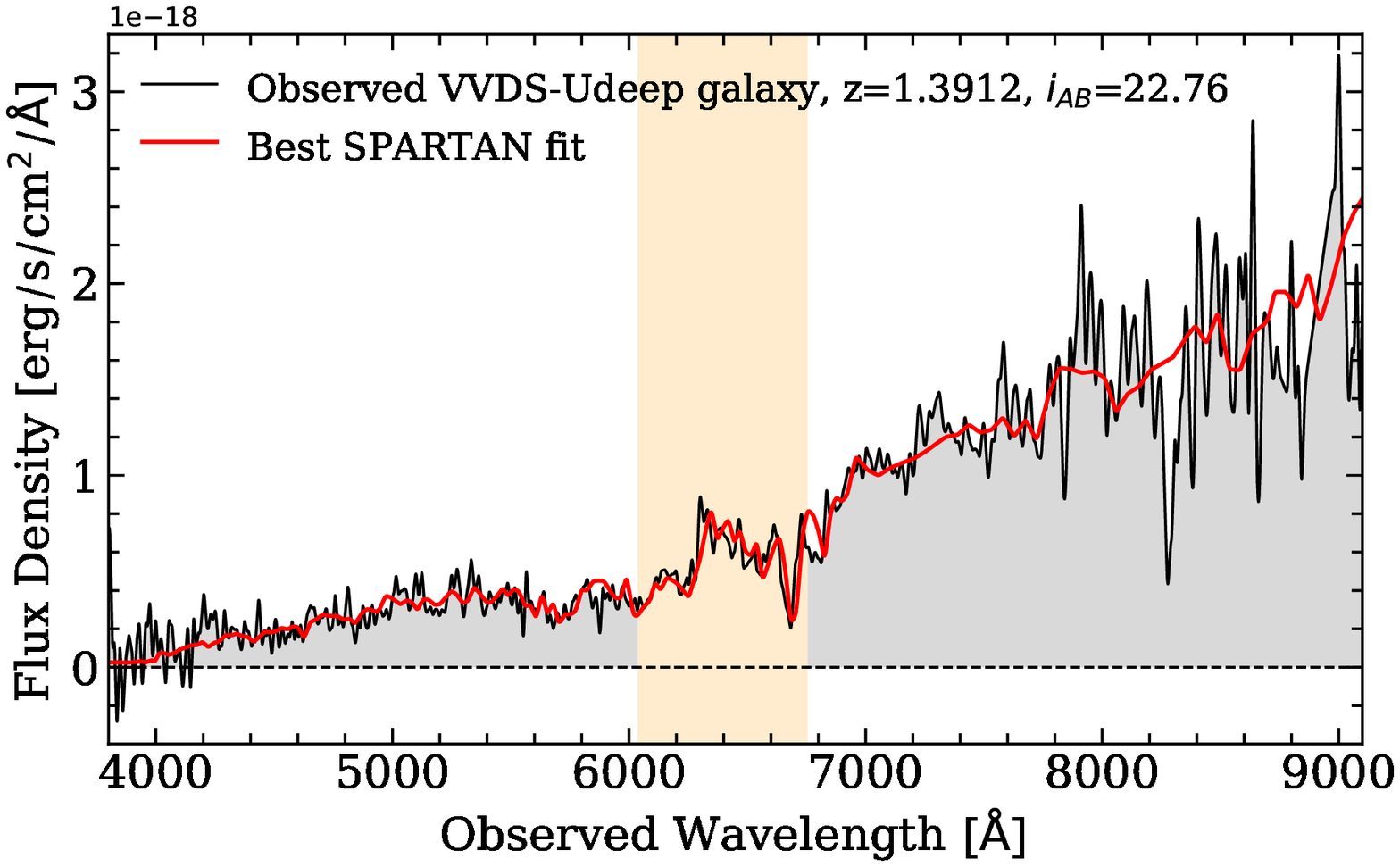}
\includegraphics[width=8.cm]{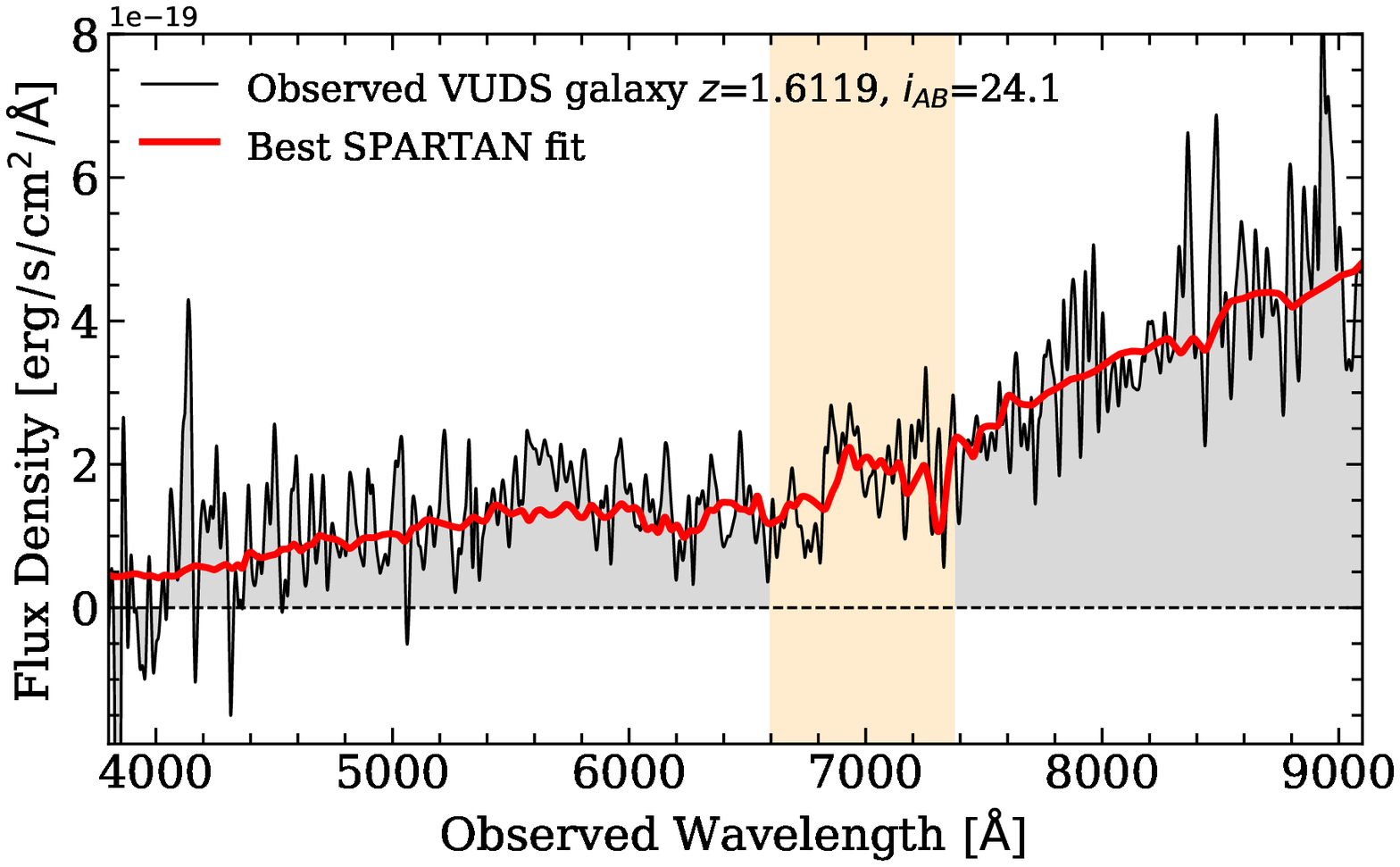}
\caption{Example of selected Mg$_{UV}$-selected galaxies at three different redshifts. For each galaxy the black line shows the observed spectrum, the red line the best fit from SPARTAN. We show with a light yellow strip the location of the Mg$_{UV}$ index. When available, we also indicate the position of the D$_{4000}$ break with a blue vertical strip. The two bottom plots contain galaxies with an assigned redshift quality flag of 1 and would have therefore been missed if low-redshift flag galaxies had been taken out of the selection from the beginning.}
\label{Spectral_fit}
\end{figure}

\begin{figure}[h!]
\centering
\includegraphics[width=0.5\textwidth]{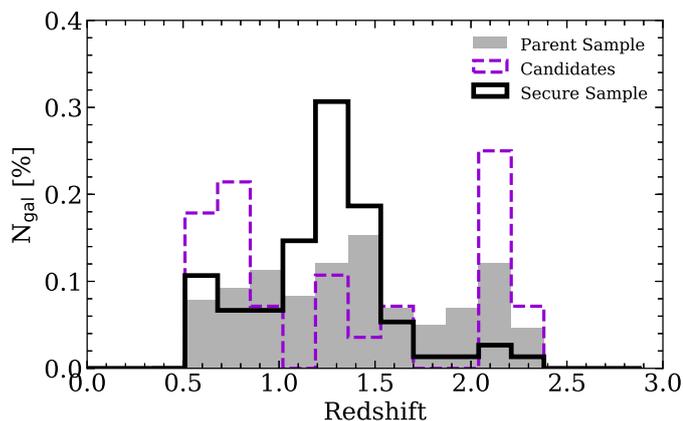}
\caption{Redshift distributions of our selected Mg$_{UV}$-selected galaxies. The filled gray histogram represents the parent sample of 3711 galaxies. The black histogram shows the secure sample while the dashed purple histogram indicates the candidate sample.}
\label{redshift}
\end{figure}

The first subsample is a secure catalog of galaxies with clear Mg$_{UV}$ index. This sample contains 75 galaxies. The mean redshift is 1.22 with a dispersion of 0.36. The mean Mg$_{UV}$ based on the measurement on data is 1.46$\pm$0.33, while taking the measurement from the best fit templates gives a mean Mg$_{UV}$ of 1.41 with a dispersion of 0.24. Examples of galaxies in the selected sample with their associated best fit from SPARTAN are presented in Fig.\ref{Spectral_fit} and show that SPARTAN is able to reproduce very well the spectra of our objects. 
The redshift distribution of this most reliable sample is shown in Fig.\ref{redshift} (black histogram). This distribution shows that we were able to select galaxies in this sample from z$\sim$0.5 to z$\sim$2.1. The distribution shows a peak $z\sim$1.2-1.4. This peaks corresponds to a Mg$_{UV}$ between $\sim$5500\AA~and$\sim$6225\AA~in the middle of the VIMOS wavelength window. As shown in the previous section, this spectral region is that for which the flux calibration is the most precise, which makes it easier to compute spectral quantities. We observe that the high-redshift end is less populated than the low-redshift end. In this redshift regime the Mg$_{UV}$ index moves toward redder wavelengths where the sky features are the most prominent. Even if the sky residuals are corrected on average for these strong skylines (see above), uncertainties in the correction may remain large at $\lambda>8000$\AA~for individual objects. 

The second sample that we generate during our selection is a ``candidate'' sample. The noise on the data for this sample does not allow us to detect the presence of the Mg$_{UV}$ index firmly, but  the fit of the spectra indicates its presence. This candidate sample is populated by 28 galaxies. The mean redshift is 1.33 with a higher dispersion with $\sigma=0.63$. Also, the redshift distribution of this sample is shown in Fig.\ref{redshift} (purple dashed histogram). This figure shows that this sample is mainly composed of galaxies at the edges of the redshift range, which explains why they are not entering the secure sample. In terms of Mg$_{UV}$, the measurements from the data give 1.39 with a dispersion of 0.37, while the estimation from the spectral fit gives Mg$_{UV}$=1.16 with a dispersion of 0.13. In the next section we use both samples together and separately to verify whether the candidate sample is consistent with the main sample. 
 
\begin{figure}[h!]
\centering
\includegraphics[width=0.5\textwidth]{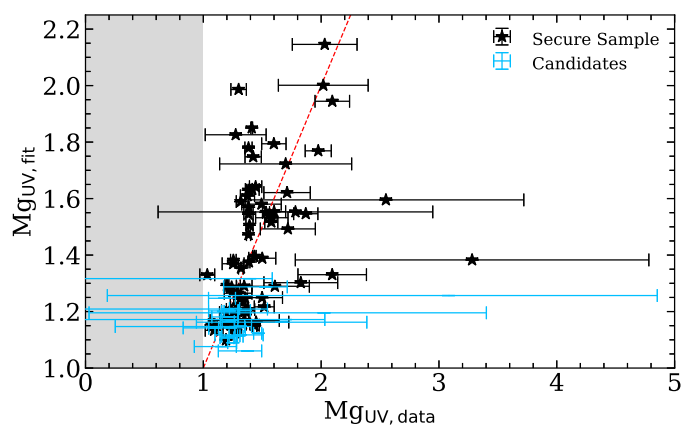}
\caption{Comparison of the Mg$_{UV}$ strength computed from the data and from the best spectral fit. The gray region represent the forbidden area with no detection of Mg$_{UV, data}$ in the data. The red dashed line shows the 1:1 relation.}
\label{mguvdatafit}
\end{figure}

Joining the two samples together leads to a sample composed of 27 galaxies from the VVDS Deep survey (24 secure and 3 candidates), 37 galaxies from the VVDS Ultra-Deep sample (29 plus 8), and 39 objects from VUDS (22 plus 17). We compare the estimation of the Mg$_{UV}$ strength computed directly from the data and from the best spectral fit template. When computing the Mg$_{UV}$ on the data we estimate the error on the Mg$_{UV}$ index using the error propagation method.
The results are presented in figure \ref{mguvdatafit}. In the secure sample, the median difference between the measurement in the best fit template and the data is 0.05 while in the candidate sample the difference reaches 0.22. This larger difference can be explained by the higher noise in the edge of the spectra. As noise can lead to large variation in the measurement of Mg$_{UV}$, in the rest of the paper we use the Mg$_{UV}$ measurement from the spectral fit models.

As stated in the previous section, the redshift flag was not incorporated in our selection. Post-factum we verified the redshift flag distribution of our selected galaxies. Among our galaxies, 53 have a redshift flag of 3 or 4 (the best possible flags, with a redshift probability to be correct >95\%). Among these galaxies 36 are from the secure sample. At a lower redshift flag of 2 and 9 (probability to be correct of at least 75\%)  we have 30 galaxies (26 in the secure sample). Finally, 20 galaxies have been assigned a redshift flag of 1 (probability to be correct higher than 50\%) and 13 of these are in the secure sample. These latter galaxies would have been missed if we had based our selection on galaxies with the best redshift flag (i.e.,  flag$\geq$2). We checked the reduced $\chi^{2}$ of the best fit for the different flags and did not see any noticeable difference. The redshift flag estimates are assigned by eye, and it is hard to recognize rare objects with spectral properties at variance with the average population of a survey. Future surveys will benefit from machine-based automated classification \citep{Jamal18}.

\section{Physical properties of our Mg$_{UV}$-selected galaxy sample}
\subsection{Stellar mass, star rormation rate, and dust extinction }
\label{secmasssfr}

\begin{figure*}[h!]
\centering
\includegraphics[width=6cm]{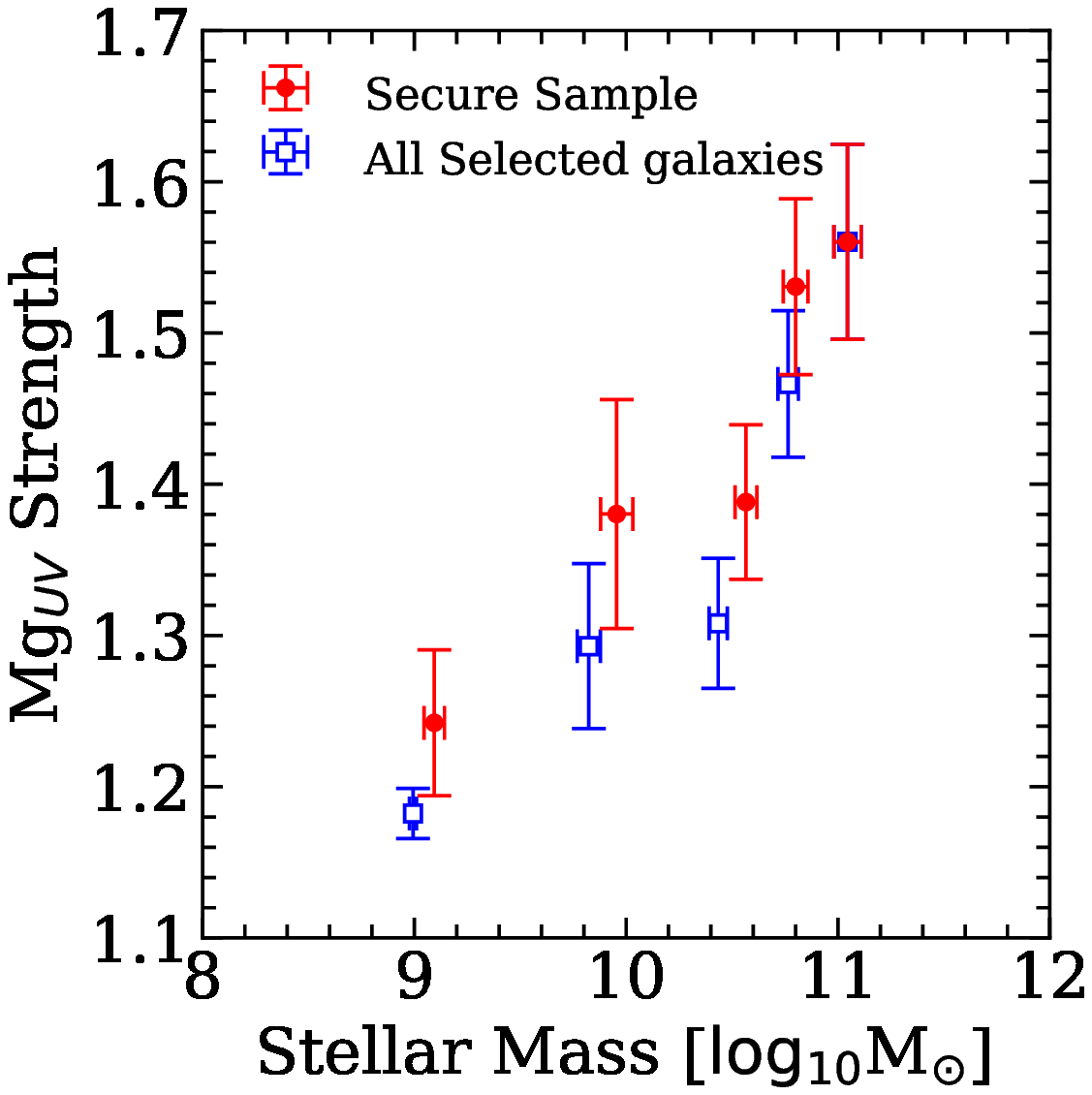}
\includegraphics[width=6cm]{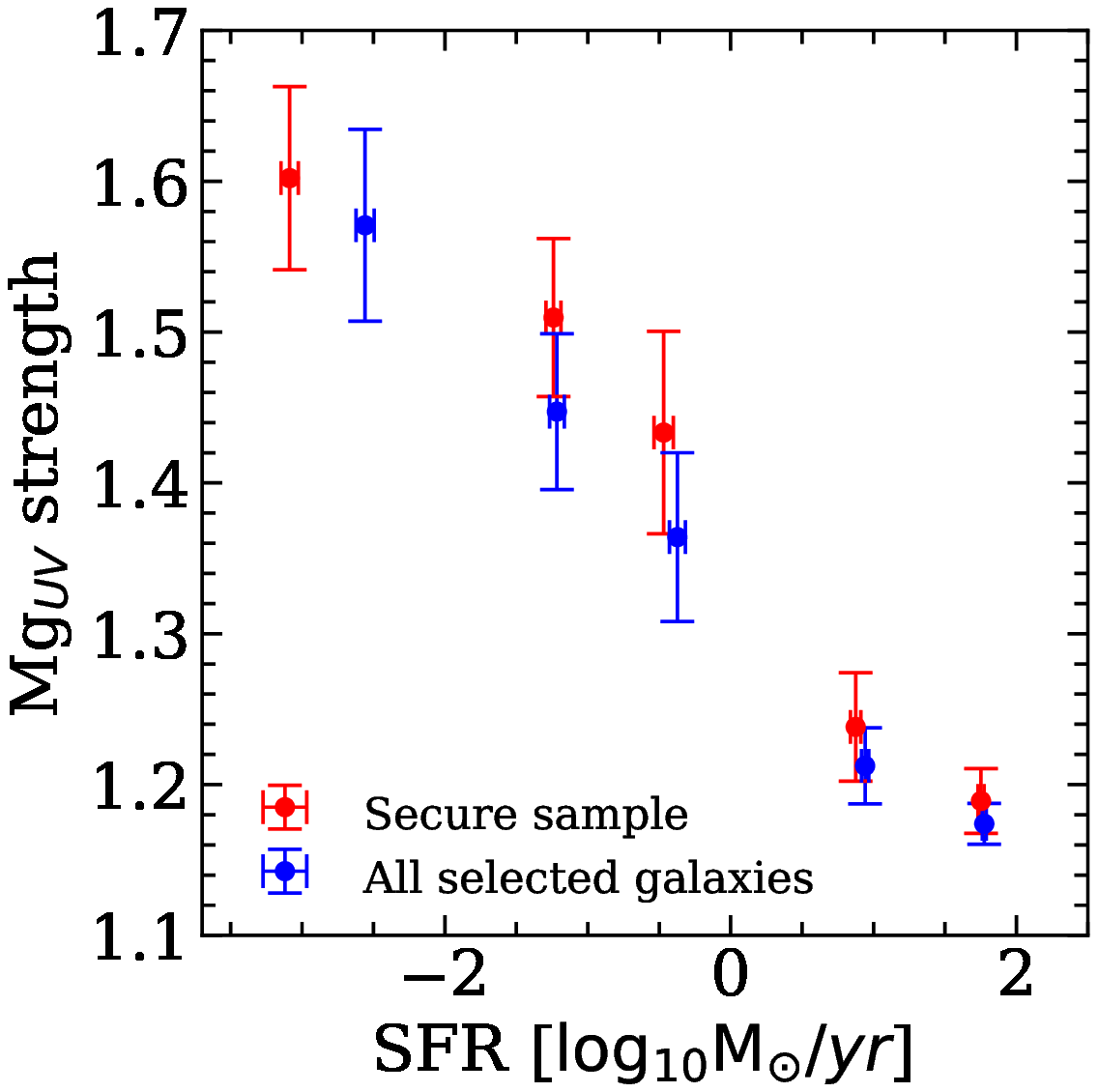}
\includegraphics[width=6cm]{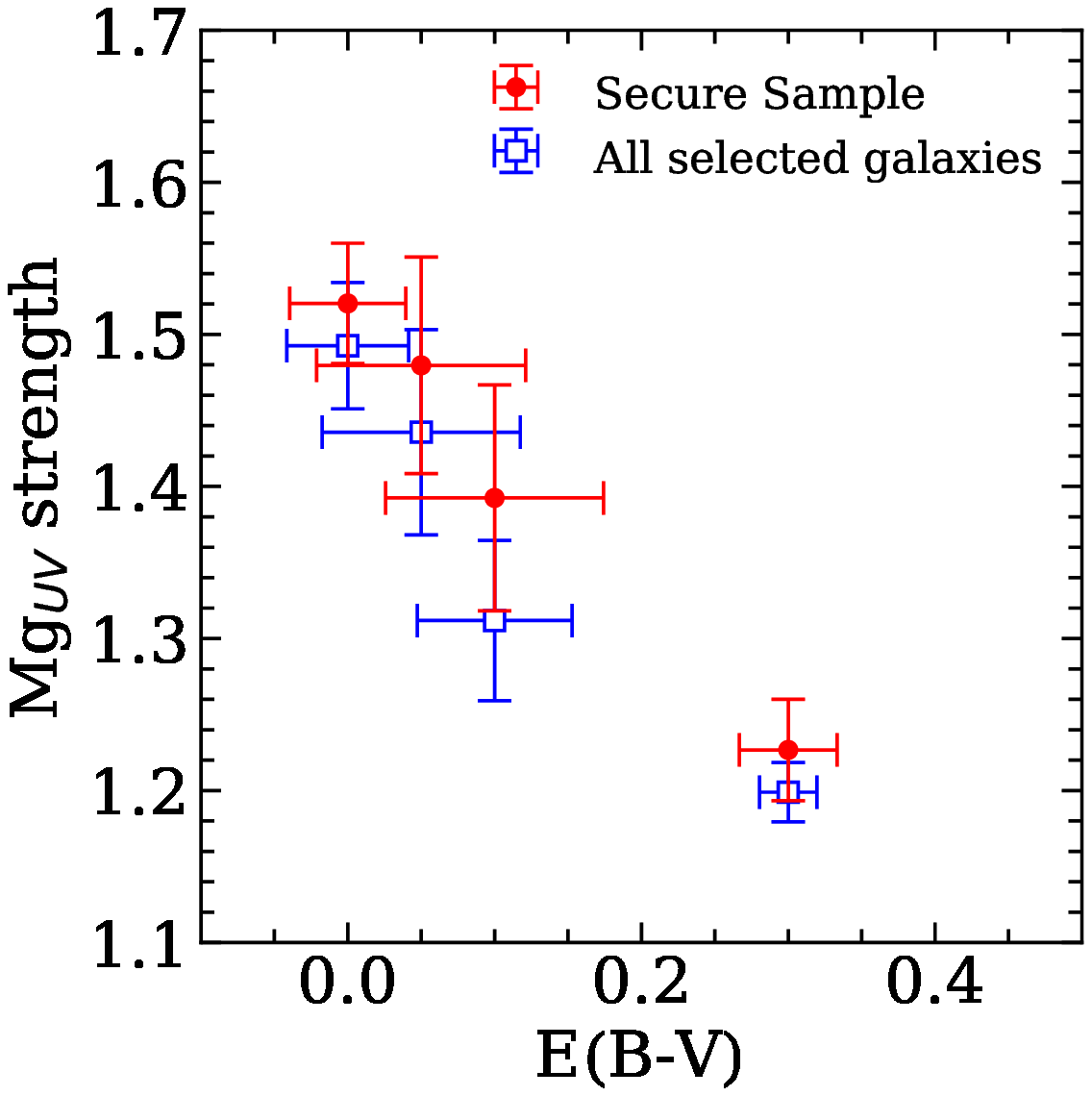}
\caption{Relationship between physical parameters computed from the SED-fitting based on photometric data only and the strength of Mg$_{UV}$ computed from the SED-fitting of the spectroscopic data only. \textit{From left to right:} Stellar mass, SFR, and dust extinction. The errors shown on each plot represent the errors on the mean. In each panel we give the evolution for the full sample of 103 galaxies (blue) and also the evolution for the secure sample only (75 galaxies, in red).}
\label{photofit}
\end{figure*}

In this section we study the properties of our selected galaxy sample as function of the strength of the Mg$_{UV}$ index. As the spectra cover a limited rest-frame spectral window, we used the photometric data alone to perform this analysis. The physical parameters were then computed using SPARTAN capabilities of fitting photometric datasets. We used the photometric bands described in Sect.\ref{Data} to perform a SED-fitting analysis of our galaxies. The library of templates used for this purpose is built making use of low resolution BC03 simple stellar population models with a \cite{Chab03} IMF. The complete parameter space we used in this fitting is described in Table \ref{Param2}.
From this fitting run we can look at the evolution of key quantities as a function of the Mg$_{UV}$ strength and we present these in Fig. \ref{photofit}. We stress that the evolution of these quantities is extracted from an independent analysis of photometric (for the physical parameters) and spectroscopic data (for the Mg$_{UV}$ index strength). It is also important to note that the binning is done on the physical parameters rather than on the Mg$_{UV}$ index itself, which explains why the absolute values of the index can differ from one plot to another.
\begin{table}[h!]
\centering
\caption{Template library used for the fitting.}
\begin{tabular}{cc}
\hline
Parameter name & Range \\ 
\hline 
\hline
SSP models  & BC03 \\ 
IMF & Chabrier (2003) \\  
Metallicity & Z>Z$_{\odot}$ \\ 
Star formation history (SFH) & Exponentially delayed\\
SFH timescale [Gyr] & 0.1 to 2.0 Gyr \\
Age [Gyr] & 0.05 up to Age$_{U}$(z)\\
Dust attenuation & Calzetti\\ 
& 0.0<E(B-V)<0.4\\
IGM & free parameter (z>1.5 only)\\
\hline 
\end{tabular} 
\label{Param2}
\end{table}

\begin{itemize}
\item \textbf{\textit{Stellar mass and SFR}}: The relationship between the stellar mass with the Mg$_{UV}$ index strength indicates that the more massive the galaxies, the stronger the Mg$_{UV}$ index. This evolution goes from Mg$_{UV}\sim$1.2 for low-mass galaxies (M$_{\star}/$M$_{\odot}\sim$9.0) to Mg$_{UV}\sim$1.55 for high-mass galaxies (M$_{\star}$/M$_{\odot}\sim$11). This confirms that galaxies exhibiting strong Mg$_{UV}$ index are in an already advanced stage of evolution with a large stellar mass.  The analysis of the evolution of the SFR indicates that this index is also particularly strong in less active galaxies. This index goes from Mg$_{UV}\sim$1.6 for the less active galaxies with SFR$\sim$0.001M$_{\odot}/$yr to Mg$_{UV}\sim$1.17 for the most active galaxies of our sample with SFR$\sim$80M$_{\odot}/$yr. This behavior means that galaxies exhibiting strong Mg$_{UV}$ are not only already massive but also have a very weak star formation activity.
\item \textbf{\textit{Dust extinction}}: The relation of the magnesium index with the dust extinction is also very clear. Galaxies with strong dust extinction E(B-V)$_{s}$=0.4 (i.e., the dustier galaxies) are also those with the weakest Mg$_{UV}$ index, at Mg$_{UV}=1.15$. Vice versa, the less dusty galaxies with  E(B-V)$_{s}\leq$0.05 present a strong magnesium break with Mg$_{UV}$=1.5. This tells us that galaxies with a strong magnesium index are the less active, most massive, and less dusty galaxies.
\end{itemize}

We also point out that the least difference in the evolution of the Mg$_{UV}$ index is minimal when considering all the selected galaxies and only the secure sample, and the same conclusions can be drawn with the inclusion of the candidate sample or without it. The only difference happens in the absolute value of the Mg$_{UV}$ in individual bins, which is always higher when removing the candidate sample.

\subsection{Galaxy ages}
\label{secage}
Galaxy age is one of the most important galaxy physical parameters and gives access to key information about galaxy evolution, such as the epoch of formation of galaxies \citep{Thomas17b}. It is also one of the more complicated parameters to estimate correctly. This is because of the numerous degeneracies that affect its estimate. The most famous are the age/metallicity \citep{Worthey99} and age-dust degeneracies \citep{Gordon97}.
Galactic age definitions and estimation methods are numerous in the literature. In 2005, D05 estimated the galactic \textit{passive age} (hereafter Age$_{passive}$) of seven identified PEGS in the Hubble Ultra-Deep Field. This age was defined as the time since the onset of passive evolution, i.e.,  when the galaxy ended its star formation period. It was estimated using the template fitting method, in which the SFH of the modeled galaxies was either exponentially declining $\tau$-models or a step-wise profile. The latter is defined as a constant SFR for a given period (a timescale $\tau$ given in gigayear) followed by a passive evolution. In each case, Age$_{passive}$ is defined as the difference between the age of the model coming from the fit and the timescale parameter $\tau$. It is equivalent to removing the star formation epoch in the measurement of the galaxy age. Another definition of the galactic age is the time elapsed since the onset of star formation, and then of the first stars of the galaxy (hereafter Age$_{onset}$). This parameter has been widely used when estimating the age from spectral indices such as D$_{4000}$ and H$\delta_{A}$ (e.g., \citealt{Siudek17}) and it is the age given directly by the template fitting method. We could argue that the time of formation of the first stars of the galaxy is hard to estimate since a single star does not define strictly the birth of a more complete and complex system that is a galaxy. For that reason, other age definitions have been used in the literature using the stellar mass estimation: the half-mass age and mass-weighted age \citep{Thomas17b}. The latter weights the age of each population of stars created in the galaxy by their own stellar mass. This ensures that more weight is given to the most important populations of stars in the galaxy. The half-mass age assumes that a system can be considered as a galaxy when half of its present stellar mass has been built up; variations of this definition can also be used, such as the quarter-mass-age.

\begin{figure}[h!]
\centering
\includegraphics[width=0.5\textwidth]{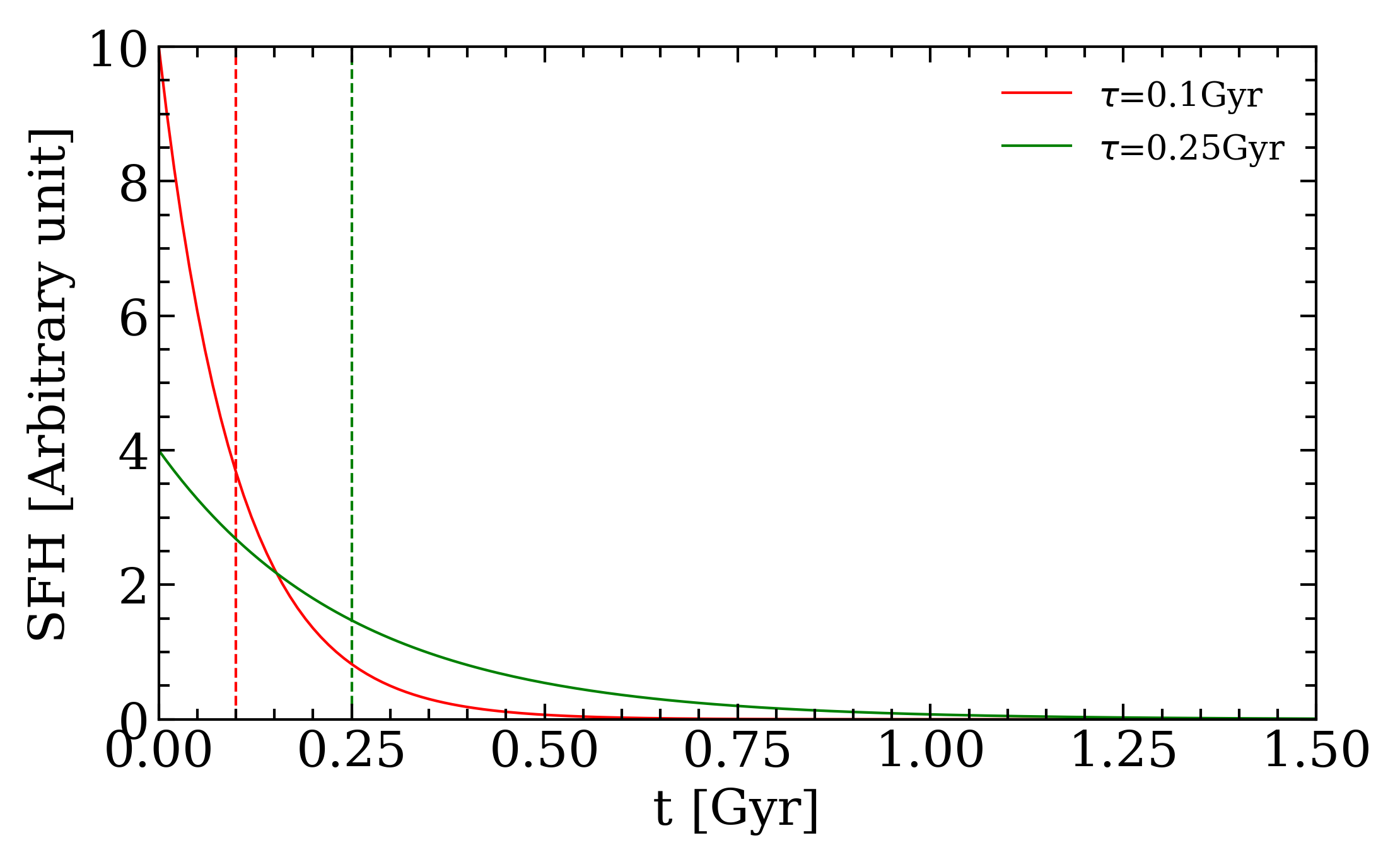}
\includegraphics[width=0.5\textwidth]{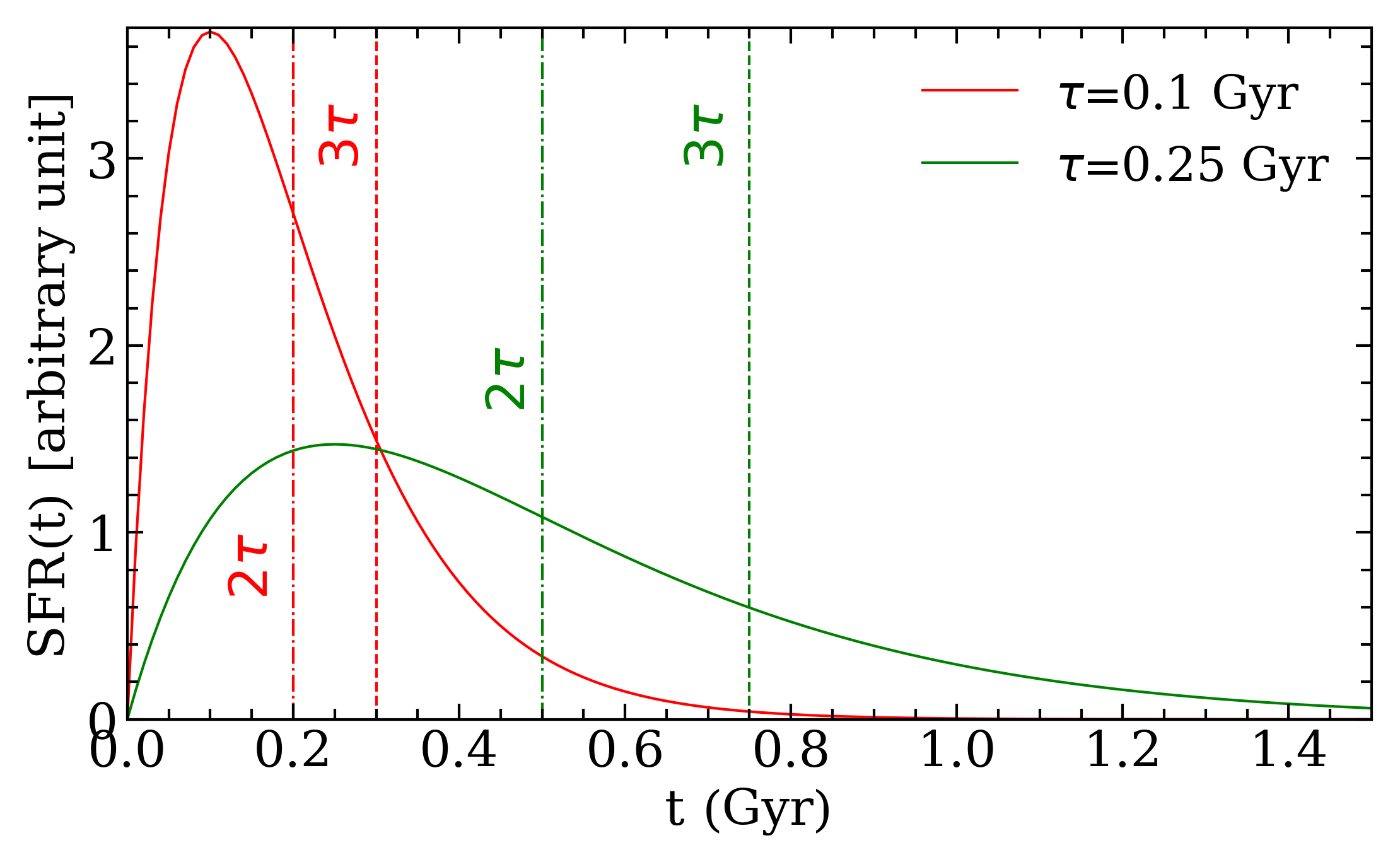}
\caption{Comparison of the Age$_{passive}$ definition between exponentially declining SFH (left panel) and exponentially delayed SFH (right panel). In each case we show the position of different timescales with dashed vertical lines.}
\label{agedef}
\end{figure}

\begin{figure}[h!]
\centering
\includegraphics[width=0.5\textwidth]{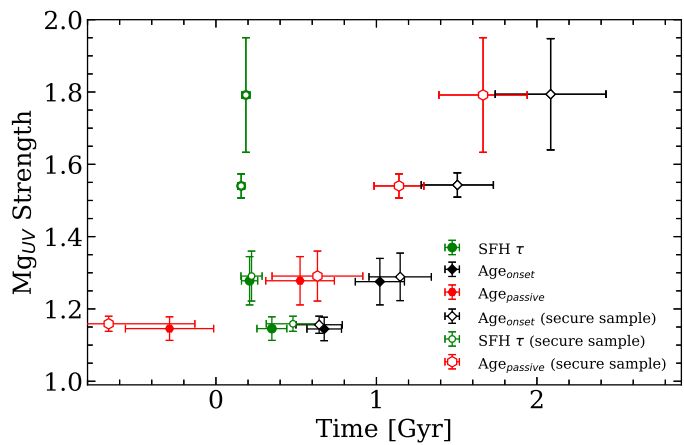}
\caption{Fitted Ages (Age$_{passive}$ in red, Age$_{onset}$ in black) and SFH timescale (in green) of our Mg$_{UV}$ selected galaxies as function of the Mg$_{UV}$ strength. The measurements from the whole sample are given with filled markers, while the measurements from the secure sample only are indicated with open markers.}
\label{agedatafit}
\end{figure}

For the purpose of this paper we computed from the SED fitting both Age$_{onset}$ and Age$_{passive}$. When using the purely declining SFH, D05 used following definition:
\begin{equation}
Age_{passive}=Age_{onset}-\tau,
\end{equation}
where $Age$ is defined as the time since the onset of star formation and $\tau$ the SFH timescale of the declining SFH since this SFH peaks at $t=0$ Gyr.  When using a different SFH, for example, the exponentially delayed SFH as in our study, this definition should be adjusted. In this prescription, the SFH peaks for $t=\tau$. Using the same definition would remove only half of the peak. Using $2\times\tau$ would allow us to remove the peak but a relatively high amount of star formation would still remain. We defined the $Age_{passive}$ as
\begin{equation}
Age_{passive}=Age_{onset}-3\times\tau.
\label{passdef}
\end{equation}
This definition seems to be a good trade-off between being too conservative (e.g., with $4\times\tau$) and not removing  the star formation epoch enough (e.g., with $2\times\tau$). It is also worth noting that using a more conservative definition (with $4\times\tau$) would not significantly change the results as the timescale parameter is on the order of 0.1 Gyr (see below). The passive galaxies would still be considered as passive galaxies even with this definition. 
With this approach we allow, by construction, Age$_{passive}$ to be negative. The results are presented in Fig. \ref{agedatafit} where we show Age$_{passive}$, Age$_{onset}$, and the SFH timescale $\tau$ for both the complete sample and the secure sample only.
The Age$_{passive}$ is always lower than Age$_{onset}$ by definition.
The joint analysis of the different ages estimates and the SFH timescale shows that galaxies with the strongest Mg$_{UV}$ are the oldest galaxies in our sample (for both age definitions). This is consistent with the fact that they are also the most massive, less active, and less dusty galaxies, as shown in Sect.\ref{secmasssfr}. For these objects, the mean Mg$_{UV}$ is $\sim$1.75. These galaxies are also those with  a very low SFH timescale with $<\tau>\sim0.16$ Gyr. This is consistent with previous studies (e.g., Daddi et al, 2004 and Thomas et al 2005) suggesting that passive galaxies have short formation timescale. It is worth mentioning that we find the timescale to be almost constant when going toward high Mg$_{UV}$ values, even though it should still decrease. This is because of the chosen binning of the value 0.1 Gyr. Therefore, any evolution smaller than this value is not visible. When the Mg$_{UV}$ weakens, the age of our galaxies decreases as well to reach a low value of Mg$_{UV}\sim$1.15. At this level, the SFH timescale reaches $<\tau>\sim0.35$ Gyr (0.40 Gyr when considering the secure sample only), while the Age$_{onset}$ decreases to $\sim$0.65 Gyr. Considering the secure sample only, it is interesting to see that Age$_{passive}$ is negative, Age$_{passive}\sim-0.66$ Gyr, which is in agreement with the definition of Age$_{passive}$. This implies that these galaxies are in the active part of the SFH. When looking also at the candidate sample this measurement is positive but at very low value with 0.05 Gyr. The fact that Age$_{onset}>\tau$ means that these galaxies have passed a peak of star formation at a very recent epoch ($\sim$250 Myr ago). The measurement of the SFR performed in the previous section confirms this analysis, where the SFR is $\sim$80 M$_{\odot}/$yr for the galaxies with the weakest Mg$_{UV}$ index. \\
This study of the physical parameters of our galaxies allows us to state that the selection of galaxies based on the Mg$_{UV}$ index allows us to select passive, massive, and old galaxies. However we point out that a weak Mg$_{UV}$ spectral signature can be present in active galaxies as well. Based on Fig.\ref{agedatafit} a threshold at Mg$_{UV}>$1.20 selects galaxies with a positive Age$_{passive}$ parameter. We apply this threshold in the rest of the paper, resulting in a sample of 66 galaxies. Out of these galaxies, nine are from the candidate sample.

\subsection{Mg$_{UV}$ galaxies in the NUVrJ and M-SFR diagrams}
\begin{figure*}[h!]
\centering
\includegraphics[width=8cm]{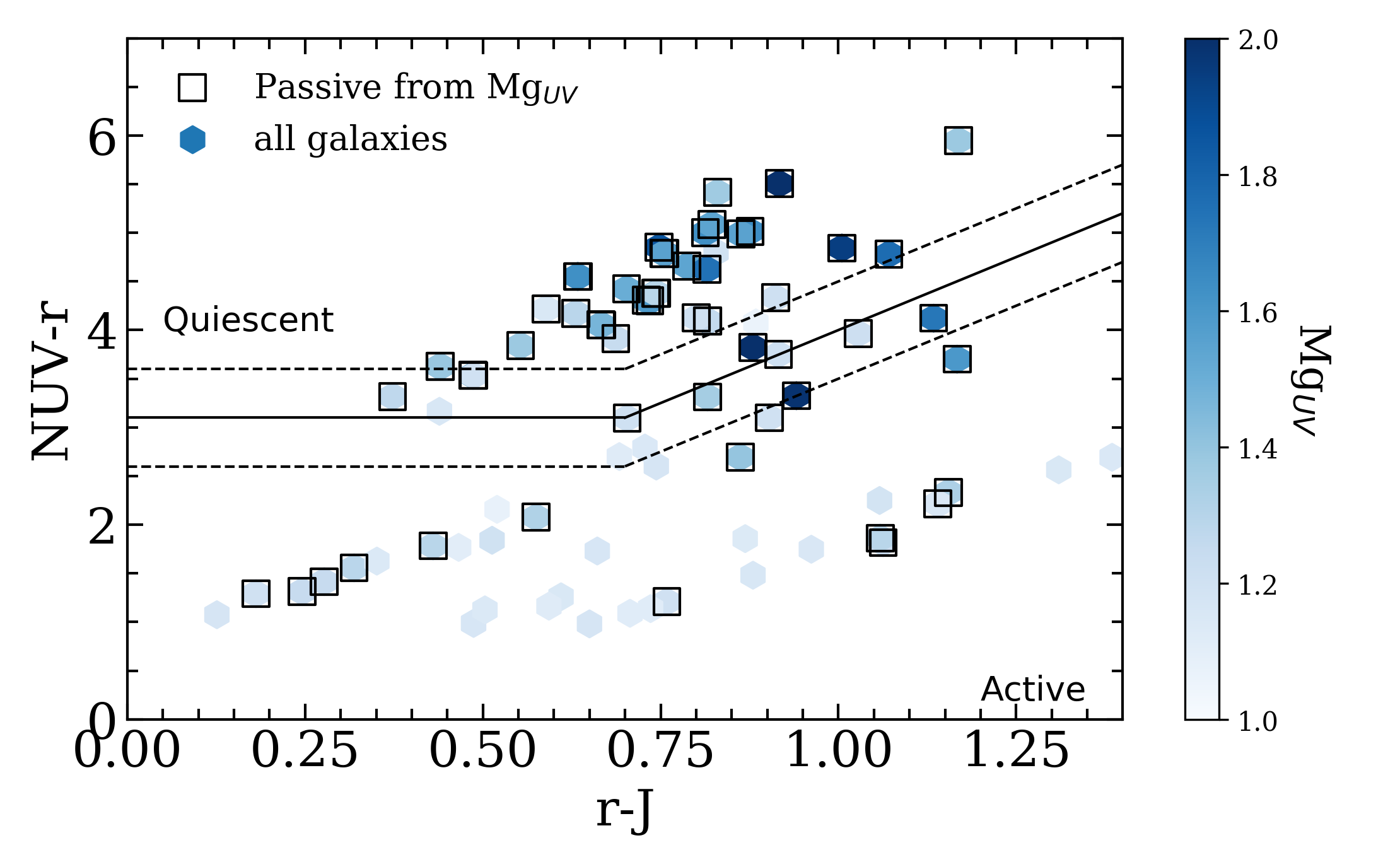}
\includegraphics[width=8cm]{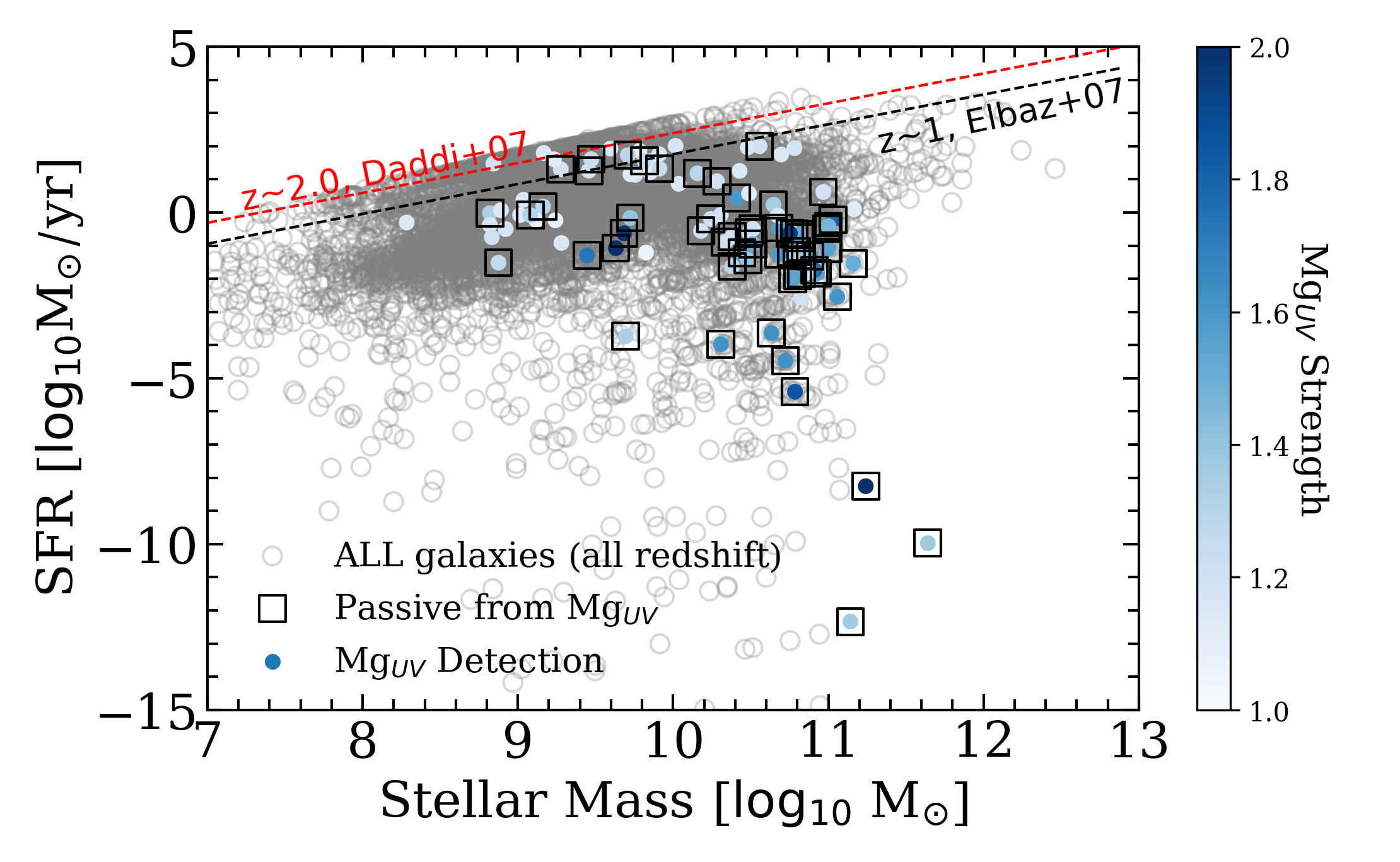}
\caption{\textit{Left:}NUVrJ color-color diagram for all our selected galaxies. The value of Mg$_{UV}$ galaxies in our sample is color-coded. We identify the galaxies selected as passive using the Mg$_{UV}$ index only by empty black squares. The passive region (black solid and dashed line) is taken from \cite{Davi17}. \textit{Right:} Position of our galaxies in the M-SFR diagram. We show in gray all the galaxies from VVDS deep and ultra deep sample (at all redshift) for comparison. We also show the main sequence fit at z$\sim$1 and 2 from \cite{elbaz07} and  \cite{daddi07}, respectively. Our selected galaxies are shown in color representing the Mg$_{UV}$ strength.}
\label{NUV}
\end{figure*}
As discussed in the introduction, passive galaxies can be selected by means of different methods: for example, color-color diagrams are widely used to extract passive galaxies from the global population. The rest-frame NUV-r-J diagram has been widely used to discriminate between active and passive galaxies \citep{Ilbert13, Davi17}. We computed the absolute magnitude of the selected galaxies in our sample to see how our galaxies behave in such a color-color selection. Our NUVrJ diagram is presented in the left plot of Fig.\ref{NUV}.
This diagram shows that two populations are clearly separated: one that clearly falls in the passive locus that is defined as \citep{Davi17} 

\begin{equation}
(NUV - r) > 3(r - J) + 1\;\;\;\mathrm{and}\;\;\;\;(NUV - r) > 3.1,
\end{equation}
and one that falls outside of this selection box.
Among our sample, 60\% of the galaxies fall into that region. The rest of our galaxies is outside of this region in the active locus. This diagram also allows us to compare our selection of passive galaxies based on the Mg$_{UV}$ only. It shows that most of the nonselected objects (active galaxies) are outside of the passive locus. We see only two galaxies in the passive locus that are not selected based on the Mg$_{UV}$ index  (5 if we take the most outer region defined by the bottom dashed line in Fig. \ref{NUV}). This seems to suggest that the Mg$_{UV}$ threshold we chose for the passive is  well suited to select passive galaxies and reduces the contamination by other galaxy types. The sSFR of these galaxies is on average <log(sSFR)>-12.4 ($\sim$3e-4 Gyr$^{-1}$) confirming their passive nature with the sSFR criterion as well. It also shows that some of our passive objects fall into the active region of the diagram, representing 20\% of these objects. The sSFR of this galaxies is on average above -9 and therefore they would not have been selected as passive using this criterion either. Nevertheless, we checked the Age$_{passive}$ parameter defined in Sect.\ref{secage} we found that it is on average 0.63 Gyr, confirming their passive nature.

It is also interesting to look at the position of our galaxies relative to average relations between SFR and $M^{*}$ for star forming galaxies in the literature  (i.e., star forming main sequence; \citealt{ elbaz07,tasca15}). The right plot of Fig.\ref{NUV} shows the position of our galaxies with respect to all the galaxies from the VVDS Deep and Ultra-Deep surveys (at all redshift) and with respect to the fit of the main sequence at z$\sim$1 and z$\sim$2. This plot shows that most of our selected passive galaxies are much below the main sequence evolution. We note that passive galaxies selected by the strength of their Mg$_{UV}$ index and that do not enter in the NUVrJ passive locus are the closest to the main sequence. This might indicate that these galaxies did not entirely terminate their evolution to a fully quiescent mode, even if they can be considered as passive from our age definition (Relation \ref{passdef}).

\section{Formation epoch and downsizing}
\label{down}

\begin{figure}[h!]
\centering
\includegraphics[width=0.5\textwidth]{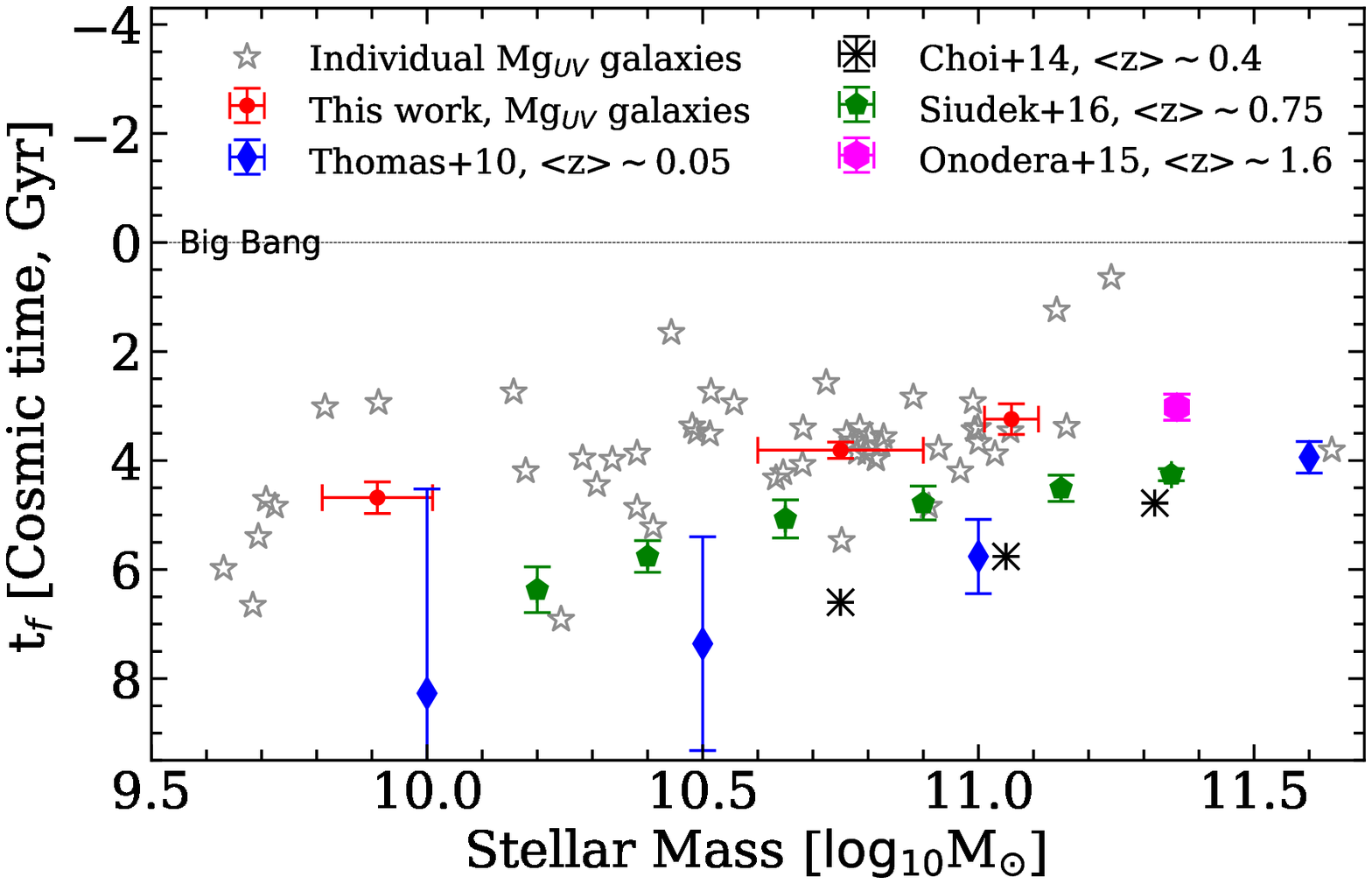}
\includegraphics[width=0.5\textwidth]{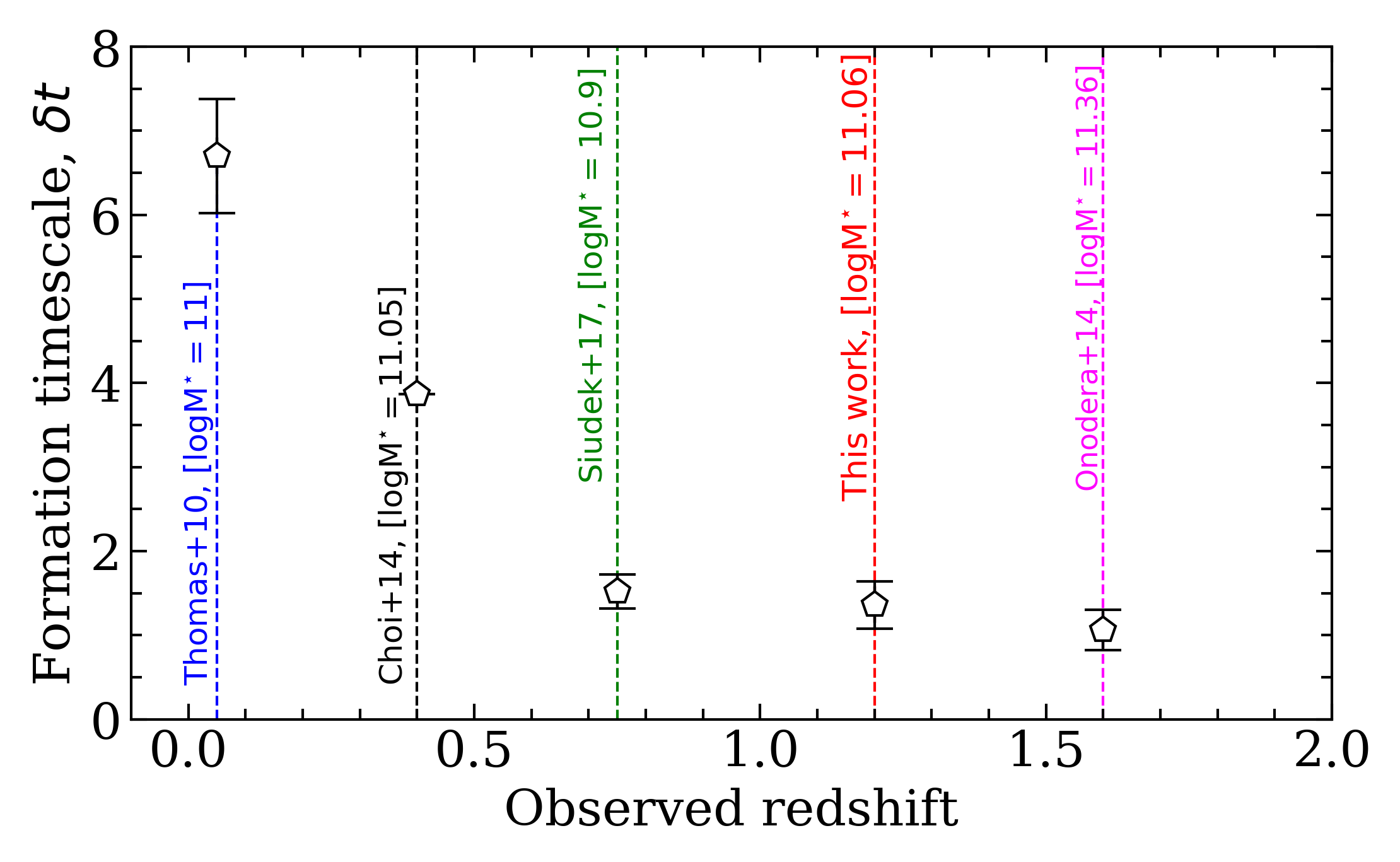}
\caption{\textit{Top:}Formation epoch (as time since the Big Bang) as a function of the stellar mass in our selected sample of magnesium galaxies (in red). We also indicate measurements from other studies in the literature with the point from \citealt{Ono15} at $<z_{obs}>\sim1.6$ (pink), measurements from \citealt{Siudek17} at $<z_{obs}>\sim0.75$ (green), points at lower redshift from \citealt{Choi14} at $<z_{obs}>\sim0.4$ (black), and from \cite{thomas10} at $<z_{obs}>\sim0.05$ (blue). \textit{Bottom:} Formation timescale defined as $\delta t = t_{obs}-t_{f}$ for galaxy of M$^{\star}\sim10^{11}$. The color coding is the same as for the top plot.}
\label{epoch}
\end{figure}

From the ages measured in the previous section we computed the formation epoch of our galaxies, $t_{f}$ (computed as the time elapsed between the Big Bang and the \textit{birth} of our galaxies). We computed this quantity using the Age$_{onset}$ definition. This definition gives access to the time of inception of the last burst of star formation that the galaxy experienced. In other words, this estimation is a lower limit on the formation epoch of the galaxy as the SFH we used contains a single burst. If other bursts took place in the evolution of the galaxies before the burst we are measuring, the age of the galaxy would be higher, and hence the formation epoch further in the past. We computed the formation time for all the galaxies in our selected sample. 

We find that the median formation epoch is at a redshift of 1.9, $\sim$3.6 Gyr after the Big Bang ($\sim$ 10 Gyr ago). As our selected sample spans a wide range of redshift, we split it into two subsamples, at $z_{obs}<1.3$ and $z_{obs}>1.3$. The low-redshift sample, with a median observed redshift of $z_{obs}\sim0.9$, has a median formation epoch of $z_{f}\sim1.8$. This is consistent with the formation redshift computed by \cite{Siudek17}. These authors used  approximately 4000 massive and passive galaxies in the VIPERS sample and found a formation redshift of $\sim$1.7 for galaxies at $z_{obs}\sim0.9$. Our highest redshift sample, which has a median observed redshift of $z_{obs}\sim1.5,$ was formed at a median epoch of $z_{f}\sim2.0$. This is in agreement with the estimation of the formation epoch of passive galaxies made by \cite{Ono15} at slightly higher redshift. In this paper the authors studied 24 early-type galaxies at $z_{obs}\sim1.6$ (less than 0.5 Gyr earlier than our galaxies in terms of cosmic time) and estimated their formation epoch at $z_{f}\sim2.0-2.5$. Our formation redshift estimates are therefore consistent with those of comparable galaxy samples in the literature.



As presented in the Introduction, numerous studies have reported a downsizing scenario of galaxy evolution. The term \textit{downsizing} has been used to describe different physical processes involving different physical quantities and \citet{Fontanot09} provides an overview of the different downsizing phenomena. In this work we are interested in \textit{archeological mass downsizing}. This term refers to different quantities and processes such as the fact that the duration of the star formation event, the burst would be shorter for these most massive galaxies \citep{thomas05}. The downsizing also refers to the fact that the most massive galaxies were formed at an earlier time \citep{Cowie96,cimatti06}. In order to study this downsizing phenomenon, we show in Fig.\ref{epoch} the dependence of the formation epoch of our galaxies as a function of the stellar mass computed in Sect.\ref{secmasssfr} for our full sample of selected galaxies at $z_{obs}\sim1.20$. 

This figure clearly shows that galaxies at high mass seems to have formed at an earlier cosmic epoch than lower mass galaxies. In the lowest observed redshift bin, galaxies with average M$_{\star}=10^{9.91}$M$_{\odot}$ are formed $\sim$4.7 Gyr after the Big Bang while galaxies with average M$_{\star}=10^{11.06}$M$_{\odot}$ are formed $\sim$3.2 Gyr after the Big Bang. Therefore, the sample of galaxies we selected supports the \textit{downsizing} phenomenon already reported in the literature. As we have explored the same mass range for all the samples (except the point from \citealt{Ono15}) we computed the average difference between the observation time and the formation time ($\delta t = t_{obs}-t_{f}$) for all the points presented in the top panel of Fig.\ref{epoch} and report the measurements in the bottom panel of Fig.\ref{epoch}. This quantity, which we call the formation timescale, tells us how long it takes to form these galaxies. For the lowest redshift sample, \cite{thomas10} found  $\delta t\sim$ 6.70 Gyr. This value decreases with increasing redshift where  the value is $\delta t\sim$3.87 Gyr for \cite{Choi14}, $\delta t\sim$1.52 Gyr for \cite{Siudek17}, $\delta t\sim$1.26 Gyr for our own sample, and $\delta t\sim$1.06 Gyr for \cite{Ono15}; the latter is at a higher stellar mass. This means that, at similar mass, galaxies observed at higher redshift took much less time to form than the galaxies at lower redshift. 

To study this result with respect to our selection we also performed the same computation for different Mg$_{UV}$ thresholds. We find that increasing the threshold on the Mg$_{UV}$ index, the points at M$_{\star}=10^{10.75}$M$_{\odot}$ and M$_{\star}=10^{11.06}$M$_{\odot}$ stay nearly at the same place, while the first point, at M$_{\star}=10^{9.91}$M$_{\odot}$, moves toward higher mass. Applying a threshold at Mg$_{UV}>$1.4, the difference reaches 0.15 in $\log_{10}$M$_{\odot}$ and 0.13 Gyr in formation time. Nevertheless, this does not change the trend we are seeing because when the threshold increases the first points goes down and accentuate the trend.\\

Finally, we observe that the t$_{f}$-M$^{\star}$ relation seems to flatten when the observed redshift is increasing. This is because the formation redshift increases with increasing observed redshift when we take the data at similar stellar masses. This is particularly visible at the low-mass end. Galaxies at M$^{\star}\sim10^{10}$ seem to have formed at $t_{f}\sim8$ Gyr when observed at $z_{obs}\sim0.05,$ while they seem to have formed at $t_{f}\sim4.6$ Gyr when observed at $z_{obs}\sim1.2$. At the high-mass end we also see the same effect but at a smaller level, which results in a flattening of the slope between the formation redshift and the stellar mass.  
The main question is whether this effect is real. It might be affected by the so-called \textit{progenitor bias} \citep{Franx96,VD01}. This bias indicates that \textit{young early-type} galaxies are included in the low-z sample but disappear in the high-z sample, creating a high-z sample  biased toward high-mass galaxies.
On the other hand, if this flattening of the $t_{f}$-M$_{\star}$ relation is not due to this bias we might be witnessing the evolution of the downsizing phenomenon that is very strong at low redshift and starts to become weaker as we observe galaxies that were formed at higher redshift. Similar studies at higher redshift are needed to test this hypothesis.

\section{Conclusions}
\label{secconc}
In this paper we studied a spectroscopic sample of galaxies selected via the strength of the Mg$_{UV}$ index. The galaxies come from three different surveys: VVDS-Deep, VVDS Ultra-Deep, and VUDS.

\begin{itemize}
\item We measure the Mg$_{UV}$ index on 3711 galaxies in these spectroscopic surveys both from the data and from the best spectral fit and select 103 galaxies with Mg$_{UV}\geq1.1$.

\item We study the evolution of several galaxy physical parameters computed from an independent photometric fitting with respect to the strength of the magnesium index Mg$_{UV}$: stellar mass, SFR, dust extinction, and age. We find a very strong correlation between each of these parameters and the specral index. We find that when its strength increases, the stellar mass of galaxies increases, their SFR decreases, the amount of dust decreases, and the galaxies becomes older. We also find that the most massive galaxies have the shortest active epoch with a SFH timescale of $\tau\sim0.1$ Gyr. This is in agreement with previous studies studying the SFHs of elliptical galaxies. We therefore conclude that the Mg$_{UV}$ index is a suitable index to identify the most quiescent (or passive) galaxies at intermediate-to-high redshift. 

\item We compute the formation redshift of our galaxies with the strongest Mg$_{UV}$ index. We find that galaxies at $<z_{obs}>\sim0.9$ were formed at an earlier epoch of $z_f \sim 1.8$ ($\sim$ 3.6 Gyr after the Big Bang) while galaxies at an higher observed redshift of  $<z_{obs}>\sim1.5$ were formed at $z_{f}\sim2$ ($\sim$3.3 Gyr after the Big Bang). These measurements are in very good agreement with previous studies carried out for passive galaxies. Moreover we compute the dependence of the formation epoch with the stellar mass (M$\star$-t$_f$ relation) and we find that the highest mass galaxies were formed at an earlier cosmic time than low-mass galaxies, supporting the downsizing scenario already reported in the literature. 

\item Finally, the M$_{\star}$-t$_f$ relation seems to be flattening as the observed redshift of the sample increases. Then the observed redshift increases the formation time of the low-mass PEGS decreases much faster than the high-mass PEGS. This effect could be partly due to the \textit{progenitor bias} already introduced in the literature. However, if this flattening is real we might be witnessing the onset of the downsizing pattern in galaxy evolution.

\end{itemize}

We can conclude that the Mg$_{UV}$ index is a good indicator of PEGs that should be considered when studying such galaxies at high redshift. It could be studied at even higher redshift when observing galaxies at NIR wavelength.

\bibliographystyle{aa}
\bibliography{final.bib}

\begin{appendix}

\section{SPARTAN tool in a nutshell}
\label{SPARTAN}
SPARTAN (SPectroscopy And photometRy fiTting tool for Astronomical aNalysis) is a SED-fitting software that is able fit both photometry and spectroscopic data. All the fits of this paper were performed using this tool and we used the capability of SPARTAN to fit a single type of data: photometry or spectroscopy. This single component fitting follows the same recipe as other codes used in the literature (e.g., GOSSIP, \citealt{Thomas17b}). For a given galaxy and a given template the $\chi^{2}$ and its associated probability are computed with

\begin{equation}
\chi^{2} = \sum_{i=1}^{N}\frac{(F_{obs,i}-A_{i}F_{syn,i})^{2}}{\sigma_{i}^{2}}; P=\exp\left[-\frac{1}{2}(\chi^{2}-\chi^{2}_{min}) \right],
\nonumber
\end{equation}

where N, F$_{obs,i}$, F$_{syn,i}$, $\sigma_i$ ,A$_i$, and $\chi^{2}_{min}$ stand for the number of observed data points, the data point itself (either a photometric band or a spectral point), the synthetic template value at the same wavelength, the observed error associated with F$_{obs,i}$, normalization factor applied to the template, and the minimum $\chi^{2}$ of the library of template, respectively. The latter is used to set the maximum of the probability distribution function (PDF) to unity and does not change the values of the estimation of the parameters nor their errors. The set of $\chi^{2}$ values are then used to create the PDF. We create from the PDF the cumulative distribution function (CDF), where the measured value of the parameter is taken where CDF(X)=0.5 and the errors on this measurement correspond to the value of the parameter for which the CDF=0.05 and 0.95.
The photometric fitting process is performed as follows. The set of synthetic templates is redshifted to the redshift of the fitted galaxies and then normalized in one predefined band. For each of the survey we studied in this paper this normalization was done in the i band. Once this normalization is done, SPARTAN convolves the normalized templates with all the photometric band passes available for the observed galaxy. Finally, the previous relations are applied to estimate the physical parameters of the observed galaxy and their associated errors. \\

When fitting spectroscopy the principle is the same except for the normalization that can be done via a different method. After redshifting the template library, SPARTAN must normalize it to the observed spectrum; SPARTAN can do this in two different ways. The first method is the same as for the photometry, i.e., it considers a photometric band pass (e.g., the i-band) and computes the magnitude from the spectrum itself. This band is then used to normalize all the templates. Nevertheless for each galaxy this band corresponds to a different rest-frame region and therefore does not treat all the galaxy in a similar manner. The second method, which we chose in this paper, is redshift-dependent and used a region of the spectrum free of emission lines. At 0.5<z<0.9, a region free of emission lines is the spectral region between 4360 and 4560\AA. When fitting a spectrum at z=0.6, the region becomes 4796-5016\AA~in the observed frame. The SPARTAN tool computes a photometric point in this region directly in the template using a box filter. This box-magnitude is used to normalize the template to the observed spectrum. At z=0.7, this region is at a redder wavelength and again a different observed region in wavelength is used to normalize the templates. This method has the advantage of being consistent from one galaxy to another. Moreover, as it is used in an emission line free region, it relies less on the emission line physics of the templates. Using the first method, the normalization is always done in a given photometric band (e.g., the i band).  For this paper we used the rest-frame regions presented in Tab.\ref{Table_norm}.
\begin{table}[h!]
\centering
\caption{Normalization region used for the spectral fit of SPARTAN}
\begin{tabular}{cccc}
\hline
Redshift range &  Normalization region [\AA] \\ 
\hline 
0.5<z<0.9   & 4360-4560\\ 
0.9<z<1.9   & 2950-3150\\
z>1.9       & 2000-2220\\
\hline 
\end{tabular} 
\label{Table_norm}
\end{table}

\section{Reproducibility}
\label{secrep}
Reproducibility in science is a crucial aspect. Sharing data and methods is as important as sharing results. We aim to address this problem in this work. In this spirit, we list all of the data-related and technique-related aspects of our work in Table \ref{Table_repro} and detail each point in the next paragraph.
\begin{table}[h!]
\centering
\caption{Summary of the reproducibility of this work}
\begin{tabular}{cccc}
\hline
 & Public & Partial & Private  \\ 
\hline 
\hline
Data VVDS-Deep  & $\surd$ & $\chi$  & $\chi$   \\ 
Data VVDS-UDeep & $\surd$ & $\chi$   & $\chi$   \\  
Data VUDS  & $\chi$ & $\surd$  & $\chi$  \\ 
SPARTAN-tool  & $\chi$ & $\surd$  & $\chi$ \\
Results & $\surd$ & $\chi$  & $\chi$ \\
Plotting tool & $\surd$ & $\chi$  & $\chi$ \\
\hline 
\end{tabular} 
\label{Table_repro}
\end{table}

\begin{itemize}
\item As presented in Sec. \ref{Data}, the VUDS sample is composed of three fields. The first data release presented in \citealt{Tasca17} is composed of all the ECDFS field and a subsample of the the COSMOS field. As such, 22 out of 27 of our VUDS objects are still private to the VUDS consortium but will be publicly released in a forthcoming paper, (Le Fevre et al, in prep). Nevertheless, we give in Appendix \ref{ap_res}, the coordinates for each of these objects. We point out that 75\% of the data we used in this paper are public and freely accessible\footnote{\url{VUDS: http://cesam.lam.fr/vuds/DR1/}}\footnote{\url{VVDS: http://cesam.lam.fr/vvds}}.
\item The SPARTAN tool is available on GITHUB\footnote{https://astrom-tom.github.io/SPARTAN/build/html/index.html} and comes with all the inputs needed to make the code run. The version released at this moment allows for a separate fit on the photometry and spectroscopy, as used in this paper. The final version will be presented in a paper in preparation (Thomas et al, in prep). We do not provide the full fit of the 3600 galaxies mainly for reasons of disk space, but the public version of SPARTAN comes with all the input to reproduce the results.
\item The IGM models used for this paper are publicly available in Zenodo \citep{igmmodels}.
\item All the measurements and results are available in Tab. \ref{Table_resudeep}, \ref{Table_resvuds} , and \ref{Table_resdeep} of the Appendix Sect.\ref{ap_res}.
\item In addition, the main python packages used during this work are public: catalog query module \textit{catscii} (v1.2, \citealt{catscii}), catalog matching algorithm \textit{catmatch} (v1.3 \citealt{catmatch}), our fits display library \textit{dfitspy} (v19.3.4, \citealt{dfitspy}), and our plotting tool, \textit{Photon} (v0.3.2, \citealt{photon}). These packages are all available in the main python package index repository (pypi).
\end{itemize}

\section{Result table}
\label{ap_res}
We provide in this appendix all the measurements performed in this paper for our 103 selected galaxies. They are given in Tables.\ref{Table_resudeep}, \ref{Table_resvuds}, and \ref{Table_resdeep}.

\begin{sidewaystable*}[h!]
\centering
\caption{Measurements and physical parameters of our galaxies from the VVDS-Ultra Deep sample. For each galaxy we provide identification, coordinates ($\alpha,\delta$), redshift, and redshift flag (z$_{flag}$; see \cite{OLF15} for more details);  measurements of Mg$_{UV}$ from data and best spectral fit; stellar masses (M$_{\star}$) and SFR (both given in $\log_{10}$ scales); the dust extinction (E(B-V)$_{s}$); Age$_{onset}$ and Age$_{passive}$ (both in Gyr); and D$_{4000}$ indexes and formation epoch (in cosmic time and in gigayearr since the Big Bang). We also give the sample ID (S: secure, C: candidate). Some galaxies appear with a value of -99.9 for the physical parameters. This means that they could not be fitted owing  to a lack of photometric points.}
\begin{tabular}{cccccccccccccccc}
\hline
ID & $\alpha$& $\delta$& redshift &z$_{flag}$ & Mg$_{UV,d}$ &  Mg$_{UV,f}$ & M$_{\star}$  & SFR & E(B-V)$_{s}$ & Age$_{onset}$ & Age$_{pass}$ & $t_f$ &D$_{4000}$ & Cat & \\ 
\hline 
\hline 
910265695       & 36.8548       & -4.3850 & 0.5116      & 3     & 1.034 & 1.330   & 8.820 & -0.021        & 0.350 & 0.719 & 0.619 & 8.17  & 1.36  & S\\ 
910378108       & 36.7234       & -4.1341 & 0.5163      & 4     & 1.100 & 1.141   & 8.283 & -0.304        & 0.150 & 0.453 & 0.353 & 8.40  & 1.34  & C\\ 
910274705       & 36.8722       & -4.3647 & 0.5249      & 3     & 1.071 & 1.154   & 8.891 & 0.070 & 0.300 & 0.806 & 0.706 & 7.99  & 1.32  & S\\ 
910359082       & 36.8471   & -4.1759 & 0.5556  & 3     & 1.092 & 1.134 & 9.014   & -0.076        & 0.200 & 1.610 & 1.210 & 6.97  & 1.40  & S\\ 
910360899       & 36.6654       & -4.1718 & 0.559       & 2     & 1.699 & 1.723   & 9.446 & -1.295        & 0.100 & 1.610 & 1.410 & 6.94  & 1.96  & S\\ 
910342467       & 36.6655       & -4.2113 & 0.5633      & 3     & 1.448 & 1.163   & 8.819 & -0.373        & 0.100 & 0.453 & 0.353 & 8.07  & 1.46  & S\\ 
910367269       & 36.7979       & -4.1581 & 0.588       & 4     & 1.216 & 1.196   & 8.832 & -0.747        & 0.050 & 0.509 & 0.409 & 7.85  & 1.57  & S\\ 
910193981       & 36.8630       & -4.5428 & 0.6125      & 3     & 1.270 & 1.211   & 10.243        & -0.197        & 0.150 & 1.280 & 0.880 & 6.91  & 1.85    & C\\ 
910202695       & 36.8607       & -4.5225 & 0.6326      & 3     & 1.316 & 1.148   & 9.280 & -0.920        & 0.050 & 2.300 & 1.900 & 5.76  & 1.65  & C\\ 
910245811       & 36.8194       & -4.4287 & 0.7833      & 2     & 1.155 & 1.076   & 9.118 & -0.232        & 0.200 & 0.453 & 0.353 & 6.73  & 1.10  & C\\ 
910195196       & 36.5937       & -4.5407 & 0.7926      & 24    & 1.383 & 1.779   & 10.910        & -1.741        & 0.000 & 2.300 & 2.100 & 4.84  & 1.99    & S\\ 
910260329       & 36.6642       & -4.3965 & 0.8612      & 3     & 1.517 & 1.217   & 10.179        & -0.558        & 0.150 & 2.600 & 2.200 & 4.19  & 1.86    & S\\ 
910296534       & 36.7009       & -4.3160 & 0.8622      & 2     & 2.030 & 2.146   & 9.631 & -1.076        & 0.100 & 0.806 & 0.706 & 5.98  & 1.86  & S\\ 
910160900       & 36.8599       & -4.6170 & 1.0259      & 3     & 1.711 & 1.621   & 10.308        & -3.974        & 0.000 & 1.610 & 1.510 & 4.44  & 1.92    & S\\ 
910197254       & 36.4191       & -4.5349 & 1.0389      & 3     & 1.347 & 1.290   & 10.381        & -1.625        & 0.050 & 1.140 & 1.040 & 4.86  & 1.91    & S\\ 
910267731       & 36.4422       & -4.3802 & 1.0537      & 3     & 1.601 & 1.554   & 10.771        & -1.998        & 0.000 & 2.300 & 2.100 & 3.64  & 1.99    & S\\ 
910215192       & 36.4405   & -4.4960 & 1.0562  & 3     & 1.382 & 1.630 & 10.633  & -3.633        & 0.000 & 1.610 & 1.510 & 4.32  & 1.92  & S\\ 
910191001       & 36.6485       & -4.5484 & 1.118       & 3     & 1.411 & 1.850   & 10.785        & -5.407        & 0.000 & 2.300 & 2.200 & 3.40  & 1.98    & S\\ 
910157260       & 36.8575       & -4.6239 & 1.1256      & 2     & 1.230 & 1.208   & 10.381        & -0.728        & 0.100 & 1.800 & 1.600 & 3.87  & 1.47    & S\\ 
910174325       & 36.8734       & -4.5860 & 1.2703      & 3     & 1.415 & 1.632   & 10.724        & -4.472        & 0.000 & 2.600 & 2.400 & 2.57  & 2.04    & S\\ 
910216226       & 36.5994       & -4.4930 & 1.2851      & 3     & 1.376 & 1.207   & 10.513        & -0.494        & 0.100 & 1.610 & 1.410 & 3.51  & 1.39    & S\\ 
910191575       & 36.7054  & -4.5464 & 1.2968   & 3     & 1.316 & 1.244 & 10.489  & -0.601        & 0.000 & 1.610 & 1.410 & 3.48  & 1.86  & S\\ 
910189008       & 36.4484       & -4.5526 & 1.3273      & 2     & 1.423 & 1.748   & 10.828        & -1.840        & 0.050 & 1.430 & 1.330 & 3.56  & 1.43    & S\\ 
910332932       & 36.7953       & -4.2322 & 1.361       & 3     & 1.201 & 1.293   & 9.708 & 1.728 & 0.350 & 0.181 & -0.019        & 4.71  & 1.16  & S\\ 
910251093       & 36.8421       & -4.4163 & 1.3868      & 3     & 1.869 & 1.546   & 10.803        & -1.899        & 0.000 & 1.280 & 1.180 & 3.54  & 2.09    & S\\ 
910334635       & 36.7042       & -4.2286 & 1.3912      & 1     & 1.392 & 1.505   & 11.160        & -1.538        & 0.000 & 1.430 & 1.330 & 3.38  & 1.41    & S\\ 
910159830       & 36.8157       & -4.6189 & 1.424       & 3     & 1.532 & 1.551   & 10.993        & -1.083        & 0.050 & 1.280 & 1.180 & 3.43  & 1.91    & S\\ 
910261703       & 36.7752       & -4.3948 & 1.4331      & 3     & 1.384 & 1.472   & 11.000        & -0.407        & 0.050 & 1.020 & 0.920 & 3.67  & 1.89    & S\\ 
910262816       & 36.8060       & -4.3916 & 1.4807      & 3     & 1.234 & 1.159   & 10.674        & -0.094        & 0.000 & 0.905 & 0.805 & 3.66  & 1.81    & S\\ 
910200844       & 36.8252       & -4.5270 & 1.5017      & 2     & 1.500 & 1.389   & 10.481        & -1.425        & 0.000 & 1.140 & 1.040 & 3.37  & 1.86    & S\\ 
910191908       & 36.8375       & -4.5460 & 1.5067      & 2     & 1.272 & 1.132   & 11.165        & 0.098 & 0.100 & 1.140 & 0.940 & 3.36  & 1.88  & S\\ 
910374671       & 36.8381       & -4.1414 & 1.54        & 2     & 1.436 & 1.150   & 10.868        & -1.803        & 0.000 & 1.280 & 1.180 & 3.13  & 1.89    & S\\ 
910293115       & 36.6150       & -4.3242 & 1.895       & 2     & 1.355 & 1.247   & 9.912 & 1.322 & 0.150 & 0.719 & 0.619 & 2.94  & 1.31  & S\\ 
910294781       & 36.6049       & -4.3203 & 2.0578      & 1     & 1.181 & 1.247   & 9.815 & 1.553 & 0.150 & 0.360 & 0.260 & 3.02  & 1.17  & C\\ 
910182978       & 36.6228       & -4.5676 & 2.1432      & 4     & 1.244 & 1.169   & 10.016        & 2.008 & 0.200 & 0.181 & 0.081 & 3.06  & 1.15  & C\\ 
910259852       & 36.7312       & -4.3975 & 2.1527      & 1     & 1.355 & 1.204   & 10.557        & 2.001 & 0.400 & 0.286 & 0.186 & 2.94  & 1.32  & C\\ 
910195919       & 36.6249       & -4.5374 & 2.2461      & 3     & 1.209 & 1.209   & -99.9 & -99.9 & -99.9 & 0.0   & -99.9 & -99.9 & -99.9 & C\\ 
\hline 
\end{tabular} 
\label{Table_resudeep}
\end{sidewaystable*}

\begin{sidewaystable*}[!htbp]
\centering
\caption{Same as \ref{Table_resudeep} but for galaxies from the VUDS survey}
\begin{tabular}{cccccccccccccccc}
\hline
ID & redshift & $\alpha$& $\delta$ &z$_{flag}$ & Mg$_{UV,d}$ &  Mg$_{UV,f}$ & M$_{\star}$  & SFR & E(B-V)$_{s}$ & Age$_{onset}$ & Age$_{pass}$ & $t_f$ &D$_{4000}$ & Cat & \\ 
\hline 
\hline 
532000150       & 0.5481        & 53.0436       & -27.7984& 4   & 1.315 & 1.117   & -99.9 & -99.9 & -99.9 & -99.9 & -99.9 & -99.9 & -99.9 & C\\ 
530041893       & 0.5765        & 53.0370       & -27.7968&1    & 3.081 & 1.257   & 8.875 & -1.515        & 0.200 & 0.905 & 0.805 & 7.53  & 1.41  & C\\ 
5170165846      & 0.6531        & 150.2343  & 2.6591& 4 & 1.300 & 1.986 & 9.684   & -0.628        & 0.100 & 1.280 & 1.080 & 6.66  & 1.82  & S\\ 
510548481       & 0.6741        & 150.1210  & 1.9350& 24        & 3.281 & 1.383   & 11.641        & -9.971        & 0.100 & 4.000 & 3.800 & 3.81  & 2.33    & S\\ 
520352454       & 0.6858        & 36.0465        & -4.8908& 1   & 1.698  & 1.172  & 8.921        & -0.501        & 0.250 & 0.509 & 0.409 & 7.23  & 1.53    & C\\ 
5101661555      & 0.726         & 150.3028  & 2.6306 &24        & 1.531 & 1.287   & 9.162 & 0.179 & 0.100 & 0.404 & 0.304 & 7.10  & 1.01  & C\\ 
530037959       & 0.7321        & 53.0534   & -27.8242  & 24    & 2.095 & 1.945   & 10.752        & -0.630        & 0.050 & 2.000 & 1.800 & 5.47  & 2.12    & S\\ 
5170122651      & 0.7377        & 150.2354  & 2.4201 &3 & 1.433 & 1.396 & 9.724   & -0.165        & 0.150 & 2.600 & 2.000 & 4.84  & 1.55  & S\\ 
5101071826      & 0.7556        & 149.7155  &   2.2758 &24      & 1.444 & 1.642   & 10.882        & -1.307        & 0.000 & 4.500 & 4.100 & 2.84  & 2.12    & S\\ 
520390517       & 0.7779        & 36.4755       & -4.3066& 3    & 1.394 & 1.317   & 9.080 & -0.079        & 0.200 & 1.140 & 0.740 & 6.08  & 1.42  & C\\ 
520374036$_B$& 0.7913   & 36.5279       & -4.3349& 3    & 1.251 & 1.090 & 9.242   & -0.237        & 0.150 & 0.509 & 0.409 & 6.64  & 1.56  & C\\ 
5101659585      & 0.823     & 150.3918   &      2.6446   & 1    & 1.827 & 1.302   & -99.900       & -99.900       & -99.900       & 0.000 & -99.9 & -99.9   & -99.9 & S\\ 
520079466       & 0.8495        & 36.0888       & -4.8508& 31   & 1.376 & 1.061   & 9.826 & -1.210        & 0.000 & 1.610 & 1.410 & 5.24  & 2.02  & C\\ 
520057061       & 0.8538        & 36.0465       & -4.8908& 4    & 2.094 & 1.331   & 9.694 & -3.734        & 0.000 & 1.430 & 1.330 & 5.40  & 1.89  & S\\ 
5100562178      & 0.8918        & 150.1987 &    1.8434 & 24     & 2.018 & 2.001   & 11.241        & -8.254        & 0.000 & 6.000 & 5.600 & 0.64  & 2.23    & S\\ 
520159600       & 0.9342        & 36.7032       & -4.7076& 2    & 1.274 & 1.147   & 10.036        & 0.865 & 0.350 & 2.300 & 0.500 & 4.15  & 1.41  & C\\ 
5101074871      & 0.9542        & 149.6587   &  2.2565   & 3    & 2.550 & 1.595   & 10.410        & 0.437 & 0.350 & 1.140 & 0.940 & 5.22  & 1.69  & S\\ 
510548050       & 0.9688        & 150.1369 &    1.9379 & 23     & 2.024 & 1.196   & 10.825        & -2.559        & 0.050 & 1.430 & 1.330 & 4.86  & 1.93    & C\\ 
510834684       & 1.02      & 149.8654 &        1.9998 &24      & 1.782 & 1.553   & 10.928        & -1.839        & 0.000 & 2.300 & 2.100 & 3.78  & 2.00    & S\\ 
520388223$_B$& 1.0508   & 36.3826       & -4.3111& 1    & 1.237 & 1.197 & 10.307  & -0.072        & 0.300 & 0.719 & 0.619 & 5.23  & 1.87  & S\\ 
520083109       & 1.102     & 36.0448   & -4.8452& 23   & 1.375 & 1.610 & 11.058  & -2.540        & 0.000 & 2.300 & 2.100 & 3.46  & 1.98  & S\\ 
530009274       & 1.1296        & 53.0535       &-28.0026& 3    & 1.557 & 1.541   & 10.443        & -1.222        & 0.000 & 4.000 & 3.600 & 1.65  & 2.05    & S\\ 
5100752385      & 1.1727        & 150.2503  &   2.0503 & 1      & 1.275 & 1.826   & -99.900       & -99.900       & -99.900       & 0.000 & -99.9 & -99.9   & -99.9 & S\\ 
520467260$_B$   & 1.2   & 36.3207       & -4.1743& 3    & 1.317 & 1.589 & 10.682  & -1.258        & 0.000 & 2.000 & 1.800 & 3.40  & 1.49  & S\\ 
520448323$_B$   & 1.3122& 36.5055       & -4.2053& 3    & 1.665 & 1.162 & 9.036   & 0.384 & 0.100 & 0.321 & 0.221 & 4.72  & 0.98  & C\\ 
5101455396      & 1.3234        & 150.1684 &    2.5389 & 22     & 1.383 & 1.545   & 10.999        & -1.095        & 0.000 & 1.610 & 1.410 & 3.40  & 1.90    & S\\ 
530054362       & 1.3408        & 53.1958       &-27.7236& 4    & 1.271 & 1.125   & 10.489        & 0.574 & 0.100 & 2.000 & 1.600 & 2.95  & 1.18  & S\\ 
530034412       & 1.5195        & 53.0059 &     -27.8436& 2     & 1.220 & 1.202   & 10.282        & 0.941 & 0.350 & 0.509 & 0.409 & 3.95  & 1.55  & C\\ 
520084083       & 1.6107        & 36.1031       & -4.8423& 2    & 1.221 & 1.165   & 10.481        & 1.959 & 0.400 & 0.286 & 0.186 & 3.96  & 1.00  & C\\ 
530045172       & 1.612     &53.0541    &-27.7772& 1    & 1.253 & 1.369 & 11.142  & -12.337       & 0.000 & 3.000 & 2.900 & 1.24  & 2.17  & S\\ 
526130207       & 1.7802        & 36.3192       & -4.2341& 1    & 1.717 & 1.493   & 10.515        & -0.957        & 0.000 & 1.140 & 1.040 & 2.74  & 1.86    & S\\ 
520118268       & 2.1226        & 36.2564       & -4.7834& 3    & 1.303 & 1.114   & 9.232 & 1.605 & 0.250 & 0.050 & -0.050        & 3.22  & 1.09  & C\\ 
520123712       & 2.1374        & 36.2261       & -4.7736& 1    & 1.467 & 1.123   & 9.725 & 1.152 & 0.150 & 0.719 & 0.519 & 2.53  & 1.28  & C\\ 
5101014292      & 2.169     & 150.1878  &       2.1607 & 3      & 1.220 & 1.181   & 10.782        & 1.942 & 0.300 & 0.360 & 0.260 & 2.84  & 1.05  & C\\ 
511040894       & 2.1863        & 149.9585 &    2.1483 &23      & 1.231 & 1.142   & 10.428        & 1.253 & 0.100 & 1.140 & 0.740 & 2.04  & 1.39  & C\\ 
5101227996      & 2.1925        & 150.1458 &    2.4125 & 2      & 1.351 & 1.168   & 9.756 & 1.134 & 0.050 & 0.286 & 0.186 & 2.88  & 1.26  & S\\ 
5101294811      & 2.2089        & 149.8866 &    2.2961 &1       & 1.607 & 1.290   & 10.157        & 1.180 & 0.150 & 0.404 & 0.304 & 2.74  & 1.37  & S\\ 
520480072$_B$& 2.2104   & 36.5049       & -4.1518& 2    & 1.281 & 1.188 & 8.842   & 1.473 & 0.150 & 0.050 & -0.150        & 3.09  & 0.99  & S\\ 
5100962101      & 2.213     & 150.4043 &        2.1674& 3       & 1.266 & 1.113   & 9.853 & 1.217 & 0.100 & 0.321 & 0.221 & 2.82  & 0.96  & C\\ 
\hline 
\end{tabular} 
\label{Table_resvuds}
\end{sidewaystable*}

\begin{sidewaystable*}[h!]
\centering
\caption{Same as \ref{Table_resudeep} but for galaxies from the VVDS-deep survey}
\begin{tabular}{cccccccccccccccc}
\hline
ID &  $\alpha$ & $\delta$ & redshift & z$_{flag}$ & Mg$_{UV,d}$ &  Mg$_{UV,f}$ & M$_{\star}$  & SFR & E(B-V)$_{s}$ & Age$_{onset}$ & Age$_{pass}$ & $t_f$ &D$_{4000}$ & Cat & \\ 
\hline 
\hline 
20099450        &  36.6058      &  -4.7890 & 1.2921     & 1     & 1.210 & 1.107   & 9.858 & 1.551 & 0.250 & 0.286 & 0.186 & 4.82  & 0.88  & C\\ 
20202638        &  36.7705      &  -4.5496 & 1.2932     & 3     & 1.255 & 1.382   & 10.814        & -0.760        & 0.000 & 1.140 & 1.040 & 3.96  & 1.87    & S\\ 
20465179        &  36.7425      &  -4.3266 & 1.315      & 2     & 1.320 & 1.354   & 10.646        & 0.233 & 0.100 & 0.806 & 0.706 & 4.23  & 1.82  & S\\ 
20260785        &  36.6414      &  -4.4270 & 1.3169     & 2     & 1.226 & 1.155   & -99.9 & -99.9 & -99.9 & 0.0   & -99.9 & -99.9 & -99.9 & S\\ 
20213373        &  37.0544      &  -4.5241 & 1.3178     & 2     & 1.206 & 1.174   & -99.9 & -99.9 & -99.9 & 0.0   & -99.9 & -99.9 & -99.9 & S\\ 
20106372        &  37.0548      &  -4.7718 & 1.3189     & 2     & 1.494 & 1.580   & -99.9 & -99.9 & -99.9 & 0.0   & -99.9 & -99.9 & -99.9 & S\\ 
20156241        &  36.7677      &  -4.6573 & 1.3233     & 2     & 1.581 & 1.532   & 10.797        & -1.178        & 0.050 & 1.140 & 1.040 & 3.87  & 1.48    & S\\ 
20467996        &  36.6398      &  -4.2816 & 1.3252     & 3     & 1.170 & 1.201   & 10.336        & -0.891        & 0.000 & 1.020 & 0.920 & 3.98  & 1.85    & C\\ 
20215973        &  36.7334      &  -4.5184 & 1.3263     & 3     & 1.232 & 1.202   & 9.276 & 1.303 & 0.250 & 0.161 & 0.061 & 4.84  & 0.89  & S\\ 
20148725        &  36.7288      &  -4.6756 & 1.3266     & 3     & 1.193 & 1.100   & -99.9 & -99.9 & -99.9 & 0.0   & -99.9 & -99.9 & -99.9 & S\\ 
20381996        &  36.3505      &  -4.1594 & 1.3279     & 1     & 1.577 & 1.517   & 10.814        & -0.671        & 0.100 & 1.020 & 0.920 & 3.97  & 1.39    & S\\ 
20215447        &  36.3945      &  -4.5199 & 1.328      & 2     & 1.387 & 1.376   & 10.824        & -1.294        & 0.000 & 1.280 & 1.180 & 3.71  & 1.47    & S\\ 
20269701        &  36.8849      &  -4.4065 & 1.3328     & 2     & 1.385 & 1.569   & 10.681        & -0.469        & 0.050 & 0.905 & 0.805 & 4.07  & 1.85    & S\\ 
20253922        &  36.9332      &  -4.4421 & 1.3345     & 2     & 1.296 & 1.152   & 10.696        & 1.748 & 0.350 & 1.430 & -0.370        & 3.54  & 1.38    & S\\ 
20385072        &  36.7800      &  -4.1539 & 1.3385     & 2     & 1.337 & 1.156   & 9.596 & 1.926 & 0.400 & 0.050 & -1.350        & 4.91  & 1.08  & S\\ 
20165698        &  36.8205      &  -4.6351 & 1.3515     & 21    & 1.975 & 1.769   & 10.990        & -0.488        & 0.000 & 2.000 & 1.800 & 2.92  & 1.61    & S\\ 
20186163        &  36.8615      &  -4.5890 & 1.3549     & 2     & 1.372 & 1.203   & 10.967        & 0.616 & 0.050 & 0.719 & 0.619 & 4.19  & 1.31  & S\\ 
20360341        &  36.8165      &  -4.2056 & 1.3592     & 2     & 1.599 & 1.794   & 11.030        & -0.218        & 0.050 & 1.020 & 0.920 & 3.88  & 1.94    & S\\ 
20256674        &  37.0463      &  -4.4358 & 1.3672     & 1     & 1.499 & 1.249   & -99.9 & -99.9 & -99.9 & 0.0   & -99.9 & -99.9 & -99.9 & S\\ 
20162681        &  36.3470      &  -4.6428 & 1.3698     & 2     & 1.247 & 1.270   & 10.780        & -0.651        & 0.000 & 1.020 & 0.920 & 3.85  & 1.34    & S\\ 
20280348        &  36.4776      &  -4.3826 & 1.3978     & 1     & 1.240 & 1.287   & 10.761        & -1.283        & 0.050 & 1.280 & 1.180 & 3.51  & 1.89    & S\\ 
20114635        &  36.8273      &  -4.7533 & 1.4572     & 2     & 1.295 & 1.265   & -99.9 & -99.9 & -99.9 & 0.0   & -99.9 & -99.9 & -99.9 & S\\ 
20130214        &  36.3206      &  -4.7192 & 1.4702     & 2     & 1.353 & 1.201   & 9.457 & 1.259 & 0.200 & 0.360 & -1.240        & 4.23  & 1.12  & S\\ 
20462585        &  36.7907      &  -4.3738 & 1.4761     & 9     & 1.231 & 1.152   & 9.166 & 1.794 & 0.250 & 0.050 & -1.950        & 4.52  & 0.99  & S\\ 
20296494        &  36.5265      &  -4.3471 & 1.5408     & 3     & 1.246 & 1.114   & 9.783 & 1.854 & 0.200 & 0.161 & 0.061 & 4.25  & 1.12  & C\\ 
20145036        &  36.5929      &  -4.6837 & 1.5413     & 1     & 1.295 & 1.223   & 9.472 & 1.605 & 0.350 & 0.128 & 0.028 & 4.28  & 1.15  & S\\ 
20147357        &  36.5588      &  -4.6787 & 1.5583     & 1     & 1.205 & 1.160   & 9.910 & 1.756 & 0.300 & 0.286 & 0.186 & 4.08  & 1.19  & S\\ 
\hline 
\end{tabular} 
\label{Table_resdeep}
\end{sidewaystable*}

\end{appendix}

\end{document}